\documentclass[prd,aps,floats,floatfix,superscriptaddress,preprintnumbers,
showpacs,eqsecnum,nofootinbib,twocolumn]{revtex4}
\usepackage{latexsym,array,theorem,mathrsfs,subfigure,bm,float}

\usepackage{psfrag}
\usepackage{amsfonts,amsmath,amssymb,latexsym,array,afterpage,
theorem,mathrsfs,bm,float,epsfig,color,graphicx,tabularx,here,multirow}

\newcommand{\nn}{\nonumber \\}
\newcommand{\bea}{\begin{eqnarray}}
\newcommand{\ena}{\end{eqnarray}}
\newcommand{\beann}{\begin{eqnarray*}}
\newcommand{\enann}{\end{eqnarray*}}

\newcommand{\lsim}{\, \mbox{\raisebox{-1.ex}
{$\stackrel{\textstyle<}{\textstyle\sim}$}}\,}
\begin{document}

\baselineskip=12pt

\title{Relativistic stars in bigravity theory}
\author{Katsuki \sc{Aoki}}
\email{katsuki-a12@gravity.phys.waseda.ac.jp}
\affiliation{
Department of Physics, Waseda University,
Shinjuku, Tokyo 169-8555, Japan
}

\author{Kei-ichi \sc{Maeda}}
\email{maeda@waseda.ac.jp}
\affiliation{
Department of Physics, Waseda University,
Shinjuku, Tokyo 169-8555, Japan
}

\author{Makoto \sc{Tanabe}}
\email{tanabe@gravity.phys.waseda.ac.jp}
\affiliation{
Department of Physics, Waseda University,
Shinjuku, Tokyo 169-8555, Japan
}

\date{\today}

\begin{abstract}
 Assuming static and spherically symmetric spacetimes 
in the ghost-free bigravity theory, 
we find a relativistic star solution, which is very close to 
that in general relativity.
The coupling constants are classified into two classes: Class [I] and Class [II].
Although the Vainshtein screening mechanism is found in the weak gravitational field for both classes,
we find that there is no regular solution beyond the critical value of the compactness in Class [I].
This implies that the maximum mass of a neutron star in Class [I] 
becomes much smaller than that in GR.
On the other hand, for the solution in Class [II],
the Vainshtein screening mechanism works well even in a relativistic star
and the result in GR is recovered.
\end{abstract}



\maketitle


\section{Introduction}
Although recent observation has confirmed the big bang scenario,
one of the biggest mysterious problems in modern cosmology is the origin of
 the current accelerating expansion of the Universe\cite{IaSN}.
One possibility to explain the accelerating expansion
is a modification of gravitational interaction on very large scale.
In general relativity (GR), it is well-known that the graviton 
is a massless spin-2 field.
The theories with a massive spin-2 field are one of the most natural extension of GR.
The non-linear ghost-free massive gravity,
often dubbed de Rham-Gabadadze-Tolley (dRGT) theory,
was proposed by de Rham et al \cite{dRGT}.
Then the dRGT massive gravity 
theory has been generalized to 
the bigravity theory by Hassan and Rosen \cite{HassanRosen},
which contains a massless spin-2 field as well as a massive spin-2 field.
The present accelerating expansion of the Universe 
can be explained in the bigravity theory
\cite{Cosmology1,Cosmology2,growth, Cosmology3,Anisotropic2, Cosmology4,Cosmology5, with_twin_matter,bigravity_dark_matter,cp_instability1,cp_instability2,cp_instability3,cp_instability4,KA_KM_RN,Mortsell}.

Since the massive gravity and the bigravity theory are 
modified by adding the mass to the graviton,
GR should be recovered by taking the massless limit.
However, the linear massive gravity, so-called Fierz-Pauli theory\cite{FP},
cannot be restored to the linearized GR even in the massless limit,
which is called the van-Dam-Veltman-Zakharov (vDVZ) discontinuity \cite{vDVZ}.
Vainshtein proposed that the vDVZ discontinuity can be  evaded by taking into account 
nonlinear mass terms
\cite{original_Vainshtein}.
Therefore, the non-linear bigravity theory may have no discontinuity in the massless limit.
In fact, GR can be recovered 
within the Vainshtein radius
in the weak-field approximation
if the coupling constants satisfy an appropriate condition 
(see later) \cite{Volkov, Babichev,Enander}.
Furthermore their analysis can be generalized into the cosmological background \cite{KA_KM_RN}
in which GR is recovered by the mechanism similar to that in the ghost condensation \cite{ghost_condensate}
 as well as by the Vainshtein mechanism.
However, in these analysis, gravitational fields are assumed to be weak.
It has not been cleared whether the Vainshtein screening mechanism
holds even in the strong gravitational field
(e.g., relativistic star and black hole). 

The black hole geometry in bigravity has been concerned,
which are classified into non-diagonal ansatz\cite{massive BH},
and bi-diagonal ansatz\cite{Volkov, Enander, Katsuragawa, Katsuragawa2, Brito}.
In the former type ansatz, there are only trivial solutions, which are the same as those in GR\footnote{There can be a hairy black hole solution in the scalar extended massive gravity\cite{BH_in_scalar_MG1,BH_in_scalar_MG2}}.
Additionally, the perturbation around the non-diagonal black hole 
is also identical to GR\cite{Kodama_Arraut, Yoshida_perturbation,babichev_nonbi}.
Hence, the massive graviton does not appear in the non-diagonal black hole.
To find a non-trivial solution, if it exists, we should assume 
both metrics can be simultaneously diagonal in same coordinate system.

There exists some special case of the bi-diagonal ansatz such that  two metrics are proportional, 
which we call a homothetic spacetime.
The solutions are also given by those in GR.
However, in this case, 
the massive graviton appears in the perturbation around the solutions.
As a result, the homothetic Schwarzschild black hole becomes 
 unstable against the radial perturbations
if the graviton mass is sufficiently small
\cite{BH_instability1,BH_instability2,BH_instability3}.
The instability of this black hole implies that there would be a hairy black hole solution as well,
and that the homothetic Schwarzschild black hole may transit to the hairy black hole.
However, the paper \cite{Brito} showed numerically that such a hairy black hole
does not exist unless the coupling constants satisfy a special condition.
One may wonder what we will find in the final stage of gravitational collapse of a compact relativistic star.
One may also ask whether there exists a maximum mass of neutron star, beyond which 
no neutron star cannot exist.

The standard picture in GR is that a star collapses to a black hole when 
the mass exceeds the maximum value.
However, in bigravity, although there exists a Newtonian star solution in the weak gravitational field,
no stable black hole solution has been found for generic coupling constants.
In order to investigate what happens when a star is compact and relativistic 
and then the gravitational interaction becomes very strong, 
we study a relativistic star in the bi-gravity theory.
A little attention has so far been paid to a relativistic star in the bigravity.
Hence, as a first step,
we analyze a star solution with a relativistic effect,
and discuss how such a relativistic star behaves 
in the limit of strong gravity.

In the text, we assume that only $g$-matter field exists
and spacetime is asymptotically flat.
We then classify the coupling constants into two classes: Class [I] and Class [II].
For Class [I],
we find an example of breaking Vainshtein screening mechanism
due to the relativistic effect.
The static star solution is found when
the pressure of the star is sufficiently small,
while the star solution disappears when the pressure is larger than a critical value.
Therefore, in Class [I], the maximum mass of the neutron star in bigravity is constrained 
stronger than one in GR.
On the other hand, there is no critical value of the pressure 
for Class [II].
The result of GR is reproduced even in the strong gravitational field.

The paper is organized as follows.
The Hassan-Rosen bigravity model
is introduced
in Sec. \ref{sec_bigravity}.
In Sec. \ref{sec_SSS_spacetime},
we derive
the basic equations in bi-diagonal ansatz
of the static and spherically symmetric spacetime.
Taking the  limit of massless graviton,
we discuss behaviours of the solutions
deep inside the Vainshtein radius in Sec. \ref{sec_massless_limit}.
We find that the existence of a neutron star solution
is restricted depending on the coupling constants.
In Sec. \ref{sec_relativistic_star},
we numerically solve the basic equations without taking the massless limit,
and confirm that the previous solutions with massless limit approximation are valid
if the Compton wave length of the graviton mass is sufficiently large compared to 
the typical radius of the star.
We summarize our results and give some remarks in Sec. \ref{summary}.
In appendix \ref{sec_weak_gravity},
we summarize the parameter constraint from the existence of a Newtonian star.
In Appendix \ref{other_sol},
introducing a cosmological constant and $f$-matter field,
we discuss solutions with asymptotically non-flat geometry.
In Appendix \ref{sec_wormhole},
we detail the case beyond the critical value of the pressure
for Class [I],
in which we find a singular behaviour.

\section{Hassan-Rosen bigravity model}
\label{sec_bigravity}
We focus on the ghost-free 
bigravity theory proposed by Hassan and Rosen \cite{HassanRosen}, whose  action is given by
\begin{eqnarray}
\!\!\!\!\!\!\!\!\!\!  S &=&\frac{1}{2 \kappa _g^2} \int d^4x \sqrt{-g}R(g)+ \frac{1}{2 \kappa _f^2}
 \int d^4x \sqrt{-f} \mathcal{R}(f) \nonumber \\
&-&
\frac{m^2}{ \kappa ^2} \int d^4x \sqrt{-g} \mathscr{U}(g,f) 
+S^{[\text{m}]}(g,f, \psi_g, \psi_f)\,,
\label{action}
\end{eqnarray}
where $g_{\mu\nu}$ and $f_{\mu\nu}$ are two dynamical metrics, and
$R(g)$ and $\mathcal{R}(f)$ are their Ricci scalars.
The parameters  $\kappa_g^2=8\pi G$ and $\kappa_f^2=8\pi \mathcal{G}$ are 
the corresponding gravitational constants, 
while $\kappa$ is defined by $\kappa^2=\kappa_g^2+\kappa_f^2$. 
We assume that the matter action $S^{[\text{m}]}$ 
is divided into two parts:
\bea
S^{[\text{m}]}(g,f, \psi_g, \psi_f)
=S_g^{[\text{m}]}(g,\psi_g)+S_f^{[\text{m}]}(f,\psi_f)
\,,
\ena
i.e.,  matter fields  $\psi_g$ and $\psi_f$ are coupled only to the $g$-metric 
and to the $f$-metric, respectively.
We call $\psi_g$ and $\psi_f$  twin matter fluids \cite{Bimond}.

 The ghost-free interaction term between the two metrics
is given by
\begin{equation}
\mathscr{U}(g,f)=\sum^4_{k=0}b_k\mathscr{U}_k(\gamma)
\,,
\end{equation}
where $\{b_k\}\,(k=0\, \mbox{-}\, 4)$ are coupling constants
and the 4$\times$4 matrix $\gamma=({\gamma^{\mu}}_{\nu})$ is 
defined by 
\begin{equation}
{\gamma^{\mu}}_{\rho}{\gamma^{\rho}}_{\nu}
=g^{\mu\rho}f_{\rho\nu}
\,, 
\label{gamma2_metric}
\end{equation}
while $\mathscr{U}_k$ are
the elementary symmetric polynomials of the eigenvalues of the matrix
 $\gamma$, defined explicitly in \cite{bigravity_dark_matter,with_twin_matter}.

Taking the variation of the action with respect to $g_{\mu\nu}$ and
$f_{\mu\nu}$, we find two sets of the Einstein equations:
\begin{align}
{G ^{\mu}}_{\nu} &=
\kappa _g^2 ( {T ^{ [\gamma ] \mu} }_{\nu} 
+ {T^{\text{[m]} \mu} }_{\nu} ) \label{g-equation}, \\
{ \mathcal{G} ^{\mu}}_{\nu} &= \kappa _f^2
( {\mathcal{T} ^{ [\gamma ] \mu} }_{\nu} 
+ {\mathcal{T}^{\text{[m]} \mu} }_{\nu} ), \label{f-equation}
\end{align}
where ${G ^{\mu}}_{\nu}$ and ${ \mathcal{G} ^{\mu}}_{\nu} $ are the Einstein 
tensors for $g_{\mu\nu}$ and $f_{\mu\nu}$, respectively. 
The $\gamma$-``energy-momentum" tensors 
${T ^{ [\gamma ] \mu} }_{\nu}$ and ${\mathcal{T} ^{ [\gamma ] \mu} }_{\nu}$
 are obtained by
the variation of the interaction term
with respect to $g_{\mu\nu}$ and $f_{\mu\nu}$, respectively, 
taking the form \cite{bigravity_dark_matter,with_twin_matter}
\begin{equation}
{T^{[\gamma] \mu}}_\nu = \frac{m^2}{\kappa^2} \left( {\tau^\mu}_\nu - {\mathscr U} {\delta^\mu}_\nu \right) \; , \quad
{{\cal T}^{[\gamma] \mu}}_\nu = - \frac{\sqrt{-g}}{\sqrt{-f}} \frac{m^2}{\kappa^2} {\tau^\mu}_\nu
\end{equation}
where 
\begin{equation}
{\cal \tau^\mu}_\nu \equiv \sum_{n=1}^4 \left( -1 \right)^{n+1} {\left( \gamma^{n} \right)^\mu}_\nu \sum_{k=0}^{4-n} b_{n+k} {\mathscr U}_k\,.
\end{equation}

The matter energy-momentum tensors 
${T ^{ [\rm m ] \mu} }_{\nu}$ and ${\mathcal{T} ^{ [\rm m ] \mu} }_{\nu}$
are given by
the variation of matter actions.
They are assumed to be 
conserved individually as
\begin{equation}
\overset{(g)}{\nabla} _{\mu}{T^{ [\text{m}] \mu} }_{\nu}=0\,,\;
\overset{(f)}{\nabla} _{\mu} {\mathcal{T} ^{ [\text{m} ] \mu} }_{\nu} =0
\,, 
\label{c1}
\end{equation}  
where $\overset{(g)}{\nabla} _{\mu}$ and $\overset{(f)}{\nabla} _{\mu}$ are 
covariant derivatives with respect to $g_{\mu\nu}$ and $f_{\mu\nu}$. 
From the contracted Bianchi identities for 
\eqref{g-equation} and \eqref{f-equation}, 
the conservation of the $\gamma$-``energy-momenta"  is
also guaranteed as
\begin{equation}
\overset{(g)}{\nabla} _{\mu}{T ^{ [\gamma] \mu} }_{\nu}=0\,,\;
\overset{(f)}{\nabla} _{\mu} {\mathcal{T} ^{ [\gamma] \mu} }_{\nu} =0
\,.
\label{c2}
\end{equation}
These equations give non-trivial constraints on solutions,
which are absent in GR.

\section{Static and spherically symmetric spacetimes}
\label{sec_SSS_spacetime}
To find a non-trivial static and spherically symmetric regular solution, 
we assume two metrics are bi-diagonal in same coordinate system.
Thus, we consider the following metric forms:
\begin{align}
ds_g^2&=-N_g^2dt^2+\frac{r_g'^2}{F_g^2}dr^2+r_g^2d\Omega^2\,, 
\label{g-metric}\\
ds_f^2&=K^2\left[-N_f^2dt^2+\frac{r_f'^2}{F_f^2}dr^2+r_f^2d\Omega^2\right]\,,
\label{f-metric}
\end{align}
where
the variables $\{N_g,F_g,r_g,N_f,F_f,r_f\}$
are functions of a radial coordinate $r$, and 
a prime denotes the derivative with respect to $r$.
The ansatz has two residual gauge freedoms: One is 
a rescaling of time coordinate ($t\rightarrow  \tilde{t}= ct$  with $c$ 
being a  constant),
and the other is 
redefinition of the radial coordinate ($r\rightarrow \tilde{r}(r)$).
The proportional constant factor $K$ is introduced just for convenience.
$K$ is one of the real roots of the quartic equation 
\begin{align}
\Lambda_g(K)&=K^2\Lambda_f(K)
\end{align}
with
\begin{align}
\Lambda_g(K)&=m^2\frac{\kappa_g^2}{\kappa^2}\,
\left(b_0+3b_1 K+3b_2 K^2+b_3 K^3\right)\,,
\nn
\Lambda_f(K)&=m^2\frac{\kappa_f^2}{\kappa^2}
\,\left(b_4+3b_3 K^{-1}+3b_2 K^{-2}+b_1K^{-3} \right)
\,. \label{eff_cc}
\end{align}
When $N_g/N_f=F_g/F_f=r_g/r_f=1$,
$g$- and $f$-spacetimes are homothetic and
the $\gamma$ energy-momentum tensors turn to be
just ``effective" cosmological terms.
In the text, we focus on asymptotically homothetic solutions, 
i.e., 
we assume the boundary condition
\begin{align}
N_g/N_f,F_g/F_g,r_g/r_f \rightarrow 1\,.
\label{boundary_cond}
\end{align}
Solutions with other asymptotic geometrical structure will be discussed in Appendix \ref{other_sol}.

We introduce new variable $\mu$ defined by
\begin{align}
\mu :=\frac{r_f}{r_g}-1
\end{align}
with $\mu>-1$,
which determines the relation between two radial coordinates $r_g$ and $r_f$.
From the boundary condition, 
$\mu$ should approach zero at infinity.

Introducing new parameters as
\begin{align}
m_g^2&:=\frac{m^2\kappa_g^2}{\kappa^2}(b_1K+2b_2K^2+b_3K^3)\,,
\\
m_f^2&:=\frac{m^2\kappa_f^2}{K^2\kappa^2}(b_1K+2b_2K^2+b_3K^3)\,,
\\
\beta_2&:=\frac{b_2K^2+b_3K^3}{b_1K+2b_2K^2+b_3K^3}\,, 
\\
\beta_3&:=\frac{b_3K^3}{b_1K+2b_2K^2+b_3K^3}\,,
\label{beta_3}
\end{align}
the Einstein equations are reduced to 
\begin{widetext}
\begin{align}
\frac{2F_g F_g'}{r_g r_g'}+\frac{F_g^2-1}{r_g^2}&=
-\kappa_g^2\rho_g-\Lambda_g+m_g^2
\left(1+2(\beta_2-1)\mu+(\beta_3-\beta_2)\mu^2
-(1+2\beta_2\mu+\beta_3\mu^2)\frac{r_f' F_g}{r_g' F_f}\right)
\,,\label{Fg_eq} \\
\frac{2 F_g^2N_g'}{r_g r_g' N_g}+\frac{F_g^2-1}{r_g^2}&=
\kappa_g^2P_g-\Lambda_g+m_g^2
\left(1+2(\beta_2-1)\mu+(\beta_3-\beta_2)\mu^2
-(1+2\beta_2\mu+\beta_3\mu^2)\frac{N_f}{N_g}\right)
\,,\label{Ng_eq}\\
\frac{2F_f F_f'}{r_f r_f'}+\frac{F_f^2-1}{r_f^2}&=
-K^2\kappa_f^2\rho_f-\Lambda_g \nn
&\quad 
+\frac{m_f^2}{(1+\mu)^2}
\left( 1+2(1+\beta_2)\mu+(1+\beta_2+\beta_3)\mu^2
-(1+2\beta_2\mu+\beta_3\mu^2)\frac{r_g' F_f}{r_f' F_g}\right)
\,,
\label{Ff_eq}\\
\frac{2F_f^2N_f'}{r_fr_f'N_f}+\frac{F_f^2-1}{r_f^2} &=
K^2\kappa_f^2P_f-\Lambda_g \nn
&\quad
+\frac{m_f^2}{(1+\mu)^2}
\left( 1+2(1+\beta_2)\mu+(1+\beta_2+\beta_3)\mu^2
-(1+2\beta_2\mu+\beta_3\mu^2)\frac{N_g}{N_f}\right)\,,
\label{Nf_eq}
\end{align}
\end{widetext}
We have two more Einstein equations, which are automatically satisfied 
since we have two Bianchi identities for $g_{\mu\nu}$ and $f_{\mu\nu}$.

In the original Lagrangian, we have six unfixed 
coupling constants $\{\kappa_f,b_i\}$,
where $m$ is not independent because it is just a normalization factor of $b_i$. 
In this paper, we use six different combinations of those constants;
$\{ m_g,m_f,\Lambda_g,K,\beta_2,\beta_3\}$,
in stead of $\{\kappa_f,b_i\}$,
because the behaviours of the solutions within the Vainshtein radius 
are characterized by $\beta_2$ and $\beta_3$
as we will see later.
The original coupling constants $\{\kappa_f,b_i\}$ are found from 
 $\{ m_g,m_f,\Lambda_g,K,\beta_2,\beta_3\}$.

The energy-momentum conservation laws of twin matters give
\begin{align}
P_g'+\frac{N_g'}{N_g}(\rho_g+P_g)&=0\,,
\label{g-matter_con}
\\
P_f'+\frac{N_f'}{N_f}(\rho_f+P_f)&=0\,,
\label{f-matter_con}
\end{align}
where we assume  that twin matters are perfect fluids.
The  energy-momentum conservation laws
of the interaction terms, which are equivalent to the Bianchi identities,
reduce to one constraint equation;
\begin{align}
&
2\left(F_g-F_f\right)
\Big( N_g(1-\beta_2+(\beta_2-\beta_3)\mu)+N_f(\beta_2+\beta_3\mu) \Big)
\nn
&+r_g(1+2\beta_2\mu+\beta_3\mu^2)
\left( \frac{F_gN_g'}{r_g'}-\frac{F_f N_f'}{r_f'} \right)=0\,.
\label{gamma_con}
\end{align}

Substituting the Einstein equations \eqref{Ng_eq} and \eqref{Nf_eq}
 into Eq. \eqref{gamma_con},
we obtain one algebraic equation:
\begin{align}
\mathcal{C}[N_g,N_f,F_g,F_f,\mu,P_g,P_f]=0\,.
\label{algebraic_eq}
\end{align}

Now we have nine variables $N_g,N_f,F_g,F_f,\mu,\rho_g, P_g, \rho_f$ and $P_f$, and six ordinary 
differential equations  \eqref{Fg_eq}-\eqref{Nf_eq}, \eqref{g-matter_con}, 
\eqref{f-matter_con}  and one algebraic equation \eqref{algebraic_eq} with 
two equations of state $P_g=P_g(\rho_g)$ and $P_f=P_f(\rho_f)$.
In order to solve those equations numerically, 
we first take the derivative of \eqref{algebraic_eq}, 
and then find  seven first-order ordinary differential equations:
\begin{align}
\frac{dX}{dr}&=\mathcal{F}_X[N_g,N_f,F_g,F_f,\mu,\rho_g,P_g,\rho_f,P_f,r]\,, 
\label{diff_eq}\\
\frac{dr_f}{dr}&=r\frac{d\mu}{dr}+\mu  +1
\nn
&=J[N_g,N_f,F_g,F_f,\mu,\rho_g,P_g,\rho_f,P_f,r]\,.
\label{Jacobian}
\end{align}
where $X=\{ N_g,N_f,F_g,F_f, P_g,P_f \}$, and $\mathcal{F}_X$ and $J$ 
do not contain any derivatives.
Here we have fixed the radial coordinate as $r_g=r$ by use of the gauge freedom.
We solve these differential equations from the center of a star ($r=0$).
In order to guarantee that the above set up gives a correct solution of our system, 
we have to impose the constraint \eqref{algebraic_eq} on the variables at the center.

Note that  the proportional factor $K$ is not necessary to be unity.
Since $K$ appears only in the form of  $K^2\rho_f$ and $K^2 P_f$, 
however, 
unless $f$ matter exists, 
the basic equations are free from the value of $K$.
In what follows, we assume that  there is no $f$-matter just for simplicity.
The $f$-matter effect on the  solution will be
 discussed in Appendix \ref{sec_f_star}.

\section{Regular compact objects :  massless limit}
\label{sec_massless_limit}
Before we present our numerical solutions, we shall discuss some analytic features 
of a compact object.
The radius of neutron star is about $10^6$cm,
while the Vainshtein radius is given typically by $10^{20}$cm 
when the Compton wave length of the graviton mass is 
the cosmological scale ($m_{\rm eff}^{-1} \sim 10^{28}$cm).
The magnitude of the interaction term, which is proportional to the 
graviton mass squared, is much smaller than the density of a neutron star.
Hence, 
the interaction term seems not to affect the structure of a neutron star.
If we ignore the interaction terms in the Einstein equations (\ref{g-equation}) and (\ref{f-equation})
(or Eqs. \eqref{Fg_eq}-\eqref{Nf_eq}), 
we just find two independent Einstein equations in GR.  
Then both spacetimes are given approximately by GR solutions,
which we can solve easily.
In bigravity theory, however, 
we have one additional non-trivial constraint equation \eqref{c2} 
(or \eqref{algebraic_eq} for a static and spherically symmetric case)
even in the massless limit.
This constraint will restrict the existence of the solutions.
In this section, we consider a compact object in this massless limit.

Note that, in this massless limit, the effective action to determine the the St\"uckelberg variable $\mu$
is given by
\begin{align}
S_{\rm eff}=-\Lambda_2^4 \int d^4x \sqrt{-g} \mathscr{U}(\mu;g_{\rm GR},f_{\rm GR})
\,,
\end{align}
where $\Lambda_2=\sqrt{m/\kappa}$, and
$g_{\rm GR}$ and $f_{\rm GR}$ are solutions in GR 
which act as like external forces to the St\"uckelberg field
\footnote{ If both metrics are Minkowski ones, this action becomes a total divergence term.
Hence it is necessary that one of them is at least a curved metric.}.
This effective action is indeed the same as the non-compact nonlinear sigma model proposed by \cite{non-compact_NLSM}.
As we will see, the massless limit approximation is valid deep inside the Vainshtein radius.
It implies that, inside the Vainshtain radius, the non-compact nonlinear sigma model
with a curved metric is obtained as the effective theory for the St\"uckelberg field.

 We analyze two models: one is a simple toy model of a relativistic star, 
 i.e., a uniform-density star, and the other is a more realistic 
  polytropic star with an appropriate equation of state for a neutron star.

\subsection{The boundary condition at ``infinity'' in the massless limit}

The boundary condition at spatial  infinity, which is outside of the Vainshtein radius, 
 is given by Eq. (\ref{boundary_cond}).
 Since the radius of a neutron star is much smaller than the Vainshtein radius, 
 there exists the weak gravity region even inside of the Vainshtein radius.
We then introduce an intermediate scale $R_I$ with $R_{\star} \ll R_I \ll R_{\rm V}$,
where $R_{\star}$ and $R_{\rm V}$ are the radius of a star and the Vainshtein radius, respectively. 
The space inside the Vainshtein radius can be divided into two regions: 
the region deep inside the Vainshtein radius ($r<R_I$) and 
the weak gravity region ($R_I<r<R_{\rm V}$), where the gravitational force is described 
by a linear gravitational potential.

From the analysis for the Vainshtein screening in the weak gravity system
 \cite{KA_KM_RN,Babichev,Enander}, 
we find that GR (or Newtonian) gravity is recovered in $r<R_{\rm V}$, 
while the homothetic solution is obtained outside the Vainshtein radius $r\gg R_{\rm V}$.
The function $\mu(r)$ changes from $-1/\sqrt{\beta_3}$ 
at small distance ($r\ll R_{\rm V}$) to $0$ at large distance ($r\gg R_{\rm V}$).
When gravity is weak,
 we find $\mu \approx -1/\sqrt{\beta_3}$ deep inside of the Vainshtein radius.
Hence we expect that $\mu \approx -1/\sqrt{\beta_3}$ at $r\approx R_I$ for a relativistic star.

We then obtain the boundary condition for a relativistic star in the massless limit as
\begin{align}
{N_g\over N_f}\rightarrow 1-{GM_g\over R_I}\approx 1 \,,\quad \mu \rightarrow -{1\over \sqrt{\beta_3}}\,,
\label{boundary_cond2}
\end{align}
as $r \rightarrow R_I$,
which we can assume $R_I\approx \infty$ 
because $R_I\gg R_{\star} $.
Note that in the massless limit, 
the Vainshtein radius turns to be infinite.

\subsection{Uniform-density star}
\label{sec_uniform}
First, we consider a uniform-density star.
Since the basic equations in the massless limit are just the Einstein equations,
we can easily solve them.
The $g$-metric of this $g$-star 
 is given by the interior and exterior Schwarzschild solutions, while 
the $f$-metric is just a Minkowski spacetime:
For the interior ($r<R_{\star}$), 
\begin{align}
F_g&=\left(1-\frac{2GM_{\star}}{R_{\star}^3}r^2\right)^{1/2}\,, 
\label{interiour_F}
\\
N_g&=N_g(0)\frac{3F_g(R_{\star})-F_g(r)}{3F_g(R_{\star})-1}\,, 
\label{interiour_N}
\\
\frac{P_g(r)}{\rho_g}&=
\frac{F_g(r)-F_g(R_{\star})}{3F_g(R_{\star})-F_g(r)}\,,
\label{interiour_P}\\
F_f&=1\,, 
\quad
N_f=N_f(0)\,, 
\end{align}
while for the exterior ($r>R_{\star}$), 
\begin{align}
F_g&=\left(1-\frac{2GM_{\star}}{r} \right)^{1/2} \,, \\
N_g&=\frac{2N_g(0)}{3F_g(R_{\star})-1}F_g(r)\,, 
\label{massless_vac1}\\
F_f&=1 \,, \quad
N_f=N_f(0)\,,
\label{massless_vac2}
\end{align}
where $R_{\star}$ and 
\begin{align}
M_{\star}:={4\pi \over 3} \rho_g R_{\star}^3  
\end{align}
are the $g$-star radius 
and  the gravitational mass, respectively.

Although we can choose $N_g(0)$ (or $N_f(0)$) 
any value by the rescaling of time coordinate,
from the boundary condition $N_g/N_f=1$ at infinity ($R_I$),
we find the ratio as
\begin{align}
\frac{N_f(0)}{N_g(0)}=\frac{2}{3F_g(R_{\star})-1}
\,.
\end{align}
Only one variable $\mu$ has not been solved.
When we find a regular solution of $\mu(r)$ for the constraint \eqref{algebraic_eq}
in the whole coordinate region ($0\leq r<\infty$)
with the boundary condition $\mu\rightarrow -1/\sqrt{\beta_3}$ as $r\rightarrow \infty$, 
we can construct a relativistic star in the bigravity theory.

First we analyze the constraint \eqref{algebraic_eq} 
at the center $r=0$ ($r_f=0$), which 
gives
\begin{align}
&
\beta_3(3P_g(0)+\rho_g)\mu_0^2+6P_g(0)(\beta_2-\beta_3)\mu_0
\nn
&~+3P_g(0)(1-2\beta_2)-\rho_g=0\,,
\label{eq_center_without_f}
\end{align}
where $\mu_0:=\mu(0)$.
This is the quadratic equation of $\mu_0$, which does not 
guarantee the existence of a real root of $\mu_0$.
In order to have a real root $\mu_0$, we have one additional  constraint as
\begin{eqnarray*}
9(\beta_2^2+\beta_3^2-\beta_3)\left({P_g(0)\over \rho_g}\right)^2
+6\beta_2\beta_3\left({P_g(0)\over \rho_g}\right)+\beta_3\geq 0\,.
\end{eqnarray*}

We then classify the coupling constants $\beta_2$ and $\beta_3$ into three cases:
(1) $\beta_2<-\sqrt{\beta_3}$, 
(2) $\beta_2>\sqrt{\beta_3}$, and
(3) $-\sqrt{\beta_3}<\beta_2<\sqrt{\beta_3}$.

In the case (1),  
the real root $\mu_0$ exists only for the restricted range of $P_g(0)/\rho_g$,
In fact, 
 there are two critical values; 
$w_{-}$ and $ w_{+} ~(w_{+}>w_{-})$,
which are defined by
\begin{align}
w_{\pm}=\frac{-\beta_2\beta_3\pm 
\sqrt{(\beta_2^2-\beta_3)(-1+\beta_3)\beta_3}}{3[\beta_2^2+(-1+\beta_3)\beta_3]}\,,
\label{critical_P}
\end{align}
 and the real root exists either if 
$P_g(0)/\rho_g<w_{-}$ or if $P_g(0)/\rho_g>w_{+}$.

On the other hand, for the case (2) and (3), 
the real root $\mu_0$ always exists for any value of $P_g(0)/\rho_g$.

Furthermore, when we take into account the finiteness of the graviton mass, 
even if it is very small, we find an additional constraint on the coupling 
constants $\{\beta_2,\beta_3\}$ from the existence of non-relativistic star 
with  asymptotically homothetic spacetime\cite{Enander} 
(see also Appendix \ref{sec_weak_gravity}).

Since the case (2) is completely excluded, we find two classes of the coupling parameters, 
which provide a relativistic star with  asymptotically homothetic spacetime, as follows: 
\footnote{This classification is also valid for a non-uniform 
star because we find the same constraint equation (\ref{eq_center_without_f}) at the center of the star.}
\\[.5em]

{\bf Class \;[I]}:~~
$\beta_2<-\sqrt{\beta_3}$ and $d_1+d_2\beta_2<0$
\\

{\bf Class \;[II]}:~~
$-\sqrt{\beta_3}\leq \beta_2 \lesssim \sqrt{\beta_3}$ and $
d_1+d_2\beta_2<0$,
\\[1em]
where $d_1$ and $d_2$ are some complicated functions of $\beta_3$, which 
are defined by (\ref{def_d1}) and (\ref{def_d2}) in Appendix \ref{sec_weak_gravity}, respectively.

\begin{figure}[tbp]
\centering
\includegraphics[width=8cm,angle=0,clip]{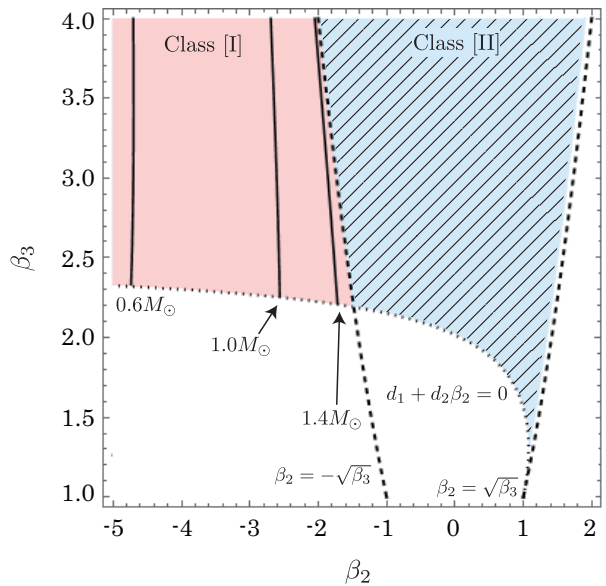}
\caption{
The constraint on the coupling constants from the existence of 
the static spherically symmetric solution in bigravity
where we assume $m_g=m_f$ and $\Lambda_g=0$.
The parameters are classified into two classes: Class [I] (the light-red region) 
and Class [II] (the hatched light-blue region).
Although there is a regular solution with Vainshtein screening mechanism
in the weak gravitational approximation for both classes,
the difference appears in the case of relativistic star.
For Class [I], the maximum mass of a neutron star is constrained
stronger than the case of GR,
while the star exists as in the case of GR for Class [II].
The contour lines of maximum mass are presented in the figure,
where the maximum mass increases as $\{\beta_2, \beta_3\}$ are close to $\beta_2=-\sqrt{\beta_3}$. 
}
\label{fig_maximum_mass}
\end{figure}

Assuming $\beta_3>1$,
 which is necessary for the existence of asymptotically homothetic solution,
we show the  ranges of Class [I] and Class [II] with 
 this constraint by the  shaded light-red region and the hatched light-blue region
in Fig.~\ref{fig_maximum_mass}, respectively.
 For its outside (the white region), there exists neither non-relativistic star
  nor relativistic one.
  
\begin{figure}[h]
\centering
\includegraphics[width=8.5cm,angle=0,clip]{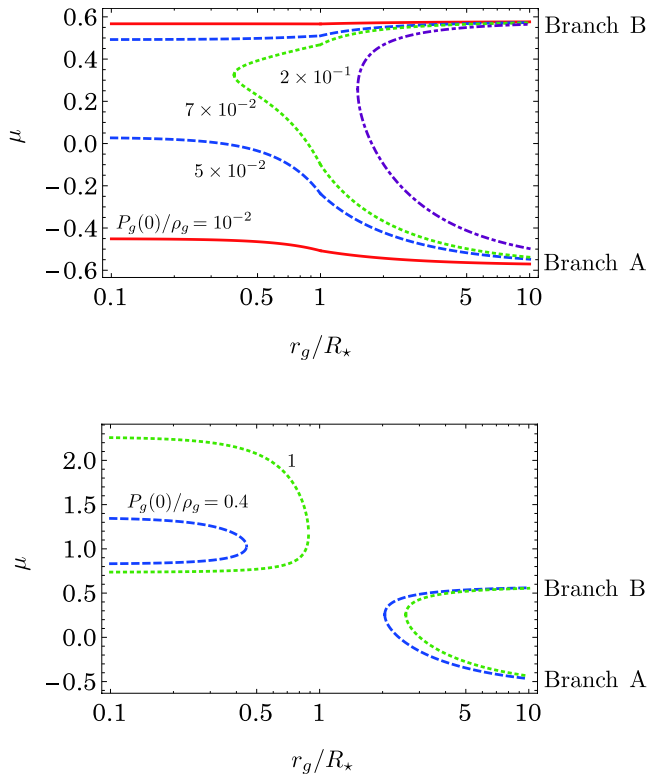}
\caption{We set $m_g=m_f,\beta_2=-3,\beta_3=3$. 
The top and the bottom figures denote  the cases of $P_g(0)/\rho_g<1/15$
and of $P_g(0)/\rho_g>1/3$, respectively.
When $1/3>P_g(0)/\rho_g>1/15$,
there is no real root of $\mu_0$.
Although
there exists a real root of $\mu_0$ for $P_g(0)/\rho_g>1/3$,
the solutions are disconnected between the region of $r\lesssim R_{\star}$ and that of $r\gg R_{\star}$.
As a result, there exist a relativistic star
for $P_g(0)/\rho_g<1/15\approx 0.06667$. 
}
\label{fig_massless_limit}
\end{figure}

\begin{figure}[h]
\centering
\includegraphics[width=8.5cm,angle=0,clip]{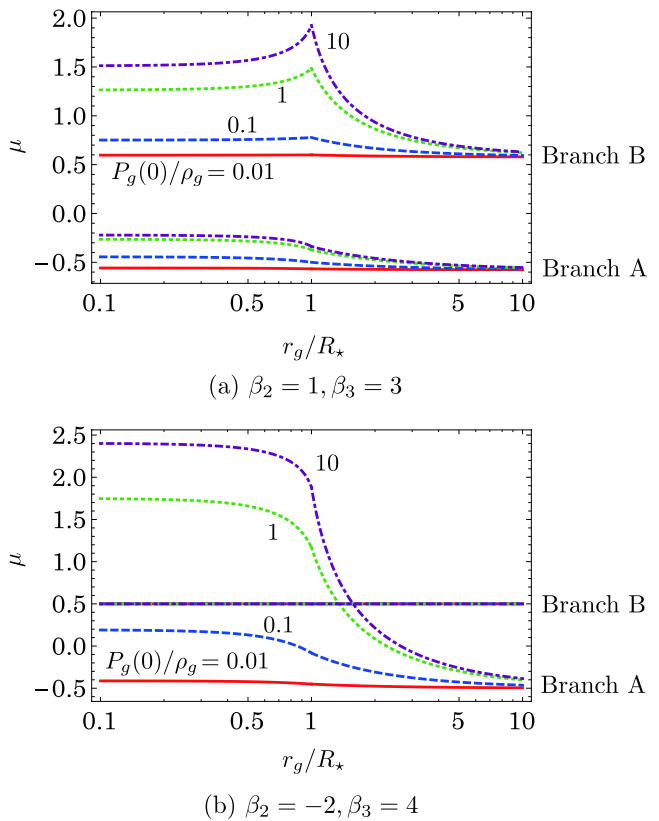}
\caption{We set $m_g=m_f$, and (a) $\beta_2=1,\beta_3=3$, 
and (b) $\beta_2=-2,\beta_3=4$.
We choose the central pressure as
$P_g(0)/\rho_g=0.01,0.1,1,10$.
In the figure (a) ($\beta_2^2-\beta_3<0$),
there are two branches, and these do not connect for a large pressure of the star.
For the figure (b) ($\beta_2^2-\beta_3=0$),
there are a non-trivial root (the branch A) as well as a trivial root   $\mu=1/\sqrt{\beta_3}$
(the branch B).
Although these two roots intersect beyond a critical pressure,
there is a regular star beyond it in the branch A.
}
\label{massless_sol3}
\end{figure}

Even if a real $\mu_0$ exists, 
we may not find a regular solution of $\mu(r)$ in the whole coordinate range ($0\leq r <\infty$)
because
the real root of \eqref{algebraic_eq}  may disappear at some finite radius.
 In  Figs. \ref{fig_massless_limit} and \ref{massless_sol3}, 
we present some examples of Class [I] and Class [II], respectively.
As the example of Class [I],  we choose
\begin{align}
m_g=m_f\,,\quad \beta_2=-3\,, \quad \beta_3=3\,,
\label{coupling_class1}
\end{align}
while for Class [II],  the parameters are chosen as
\begin{align}
m_g=m_f\,, \quad \beta_2=1\,,\quad \beta_3=3\,,
\label{coupling_class2}
\end{align}
and
\begin{align}
m_g=m_f \,, \quad \beta_2=-2\,,\quad \beta_3=4\,.
\end{align}
Note that there are two real roots for $\mu_0$.
Then we find two branches of $\mu(r)$, which we call the branch A and the branch B.
The branch A approaches a homothetic solution ($\mu\rightarrow -1/\sqrt{\beta_3}$) as
$r\rightarrow \infty$ in the massless limit,  
while the branch B ($\mu\rightarrow 1/\sqrt{\beta_3}$) does not become homothetic at infinity.

For the Class [I] example (\ref{coupling_class1}),  $\mu_0$ exists only if 
$P_g(0)/\rho_g <w_-=1/15$ (the top figure of Fig. \ref{fig_massless_limit}) or 
 $P_g(0)/\rho_g>
w_+=1/3$ (the bottom figure).
We find a regular solution for both branches if $P_g(0)/\rho_g <1/15$.
The branch A solutions provide relativistic stars with asymptotically homothetic 
spacetime, while the branch B solutions are not asymptotically flat.

For   $1/15<P_g(0)/\rho_g <1/3$, $\mu_0$ does not exist. 
We find the solution $\mu(r)$ only for the region larger than some finite radius,
and  two branches A and B are connected.
The topology of this spacetime is similar to a wormhole, but it has a curvature singularity 
at the throat (the turning point of $\mu(r)$).
For the large value of $P_g(0)/\rho_g$, the turning point appears outside of the ``star'',
which means the ``wormhole'' structure exists even for the vacuum case.
 (We should analyze the original equations without matter,
which will be done in Appendix \ref{sec_wormhole}).
Therefore, the existence of such a wormhole type solution 
may be caused by the strong gravity effect
rather than the effect of the pressure.

The wormhole throat corresponds to the point $d\mu/dr_g
= \infty$ (i.e., $dr_f/dr_g= \infty$).
When we have $dr_f/dr_g=\infty$, 
the interaction terms diverges at the point.
As a result, the contribution from the interaction term 
should not be ignored even for the case with a very small graviton mass,
and then our assumption is no longer valid at a wormhole throat.
Hence, we have to re-investigate whether 
 a relativistic star does not exist for the coupling constants of Class [I].
We shall analyze it in next section.

When $P_g(0)/\rho_g$ becomes larger, i.e., if 
$P_g(0)/\rho_g>1/3$, we again find a real $\mu_0$, 
but there exists no regular $\mu(r)$ for the whole range of $r$.
$\mu(r)$ exists in two separated regions; one is smaller than some finite radius 
($<R_\star$) and the 
other is larger than another finite radius ($>R_\star$) ,
In both regions,  two branches A and B are connected.
We find a kind of closed universe for the smaller-radius inner region,
and a kind of wormhole structure for the larger radius outer region.
Both spacetime structures contain a curvature singularity at the throats
(the turning points of $\mu(r)$).

On the other hand, for the Class [II] example,  both 
branch A and B solutions exist 
for any value of $P_g(0)$ (Fig. \ref{massless_sol3}),
and they are  not connected each other.
Hence we always find a relativistic star with asymptotically homothetic 
spacetime structure (the branch A solution).

We note that
at the boundary of Class [I] and Class [II] (i.e., $\beta_2=-\sqrt{\beta_3}$).
The trivial solution $\mu=1/\sqrt{\beta_3}$ gives the branch B.
While the branch A has a non-trivial solution
 shown in Fig. \ref{massless_sol3} (b), which 
gives 
a relativistic star for any value of $P_g(0)$.

Hence we may conclude that a relativistic star always exists a regular solution
 for the coupling constants of Class [II].
On the other hand, there
does not exist a relativistic star 
beyond a critical value of the pressure for the coupling constants 
of Class [I], i.e., for $P_g(0)/\rho_g>w_-$. 
Instead, the spacetime may turn to a wormhole geometry with a singularity 
(or a closed universe with a singularity).

The existence condition of $P_g(0)/\rho_g<w_-$ can 
be rewritten by the compactness of a star,
$
{GM_\star/ R_\star}
$.
Using the internal solution (\ref{interiour_F}) and (\ref{interiour_P}), we find
\bea
{GM_\star\over R_\star}={2{P_g(0)\over \rho_g}\left(1+2{P_g(0)\over \rho_g}\right)\over
 \left(1+3{P_g(0)\over \rho_g}\right)^2}
 \,.
\ena
Then we obtain the existence condition for Class [I] as
\bea
{GM_\star\over R_\star}<{GM_\star\over R_\star}\Big{|}_{\rm max}:={2w_-\left(1+2w_-\right)\over
 \left(1+3w_-\right)^2}
 \,.
\ena
This gives the maximum value of the compactness of a relativistic star for given coupling constants
$\beta_2$ and $\beta_3$.
Since $\beta_2$ and $\beta_3$,
 are restricted as shown in Fig. \ref{fig_maximum_mass},
we can evaluate the upper bound of the compactness for Class [I] as
\bea
{GM_\star\over R_\star}\Big{|}_{\rm ub}:=\sup_{\rm Class [I]}
 \Big{\{}{GM_\star\over R_\star}\Big{|}_{\rm max} \Big{\}}
\approx 0.23
\,,
\ena
 which is realized for $\beta_2\simeq -1.48, \beta_3\simeq 2.19$.
 
While in Class [II], any coupling constants give the same maximum value
of the compactness, that is,
\bea
{GM_\star\over R_\star}\Big{|}_{\rm max}={4\over 9}
\,,
\ena
which is obtained from the existence condition for a regular 
interior solution in GR because there is no additional constraint 
in this class.

The upper bound of the compactness in Class [I] is almost the same as 
the observed value 
(e.g., the compactness is about $0.3$ when 
a radius of a two solar mass neutron star is $10$ km,
while it is about $0.21$ for a two solar mass star with a radius of $14$ km
\cite{star_radius1,star_radius2,star_radius3}.).
In order to give a stringent constraint on the theory by observations, 
we have to analyze more realistic star, which will be discussed in the next subsection.

\subsection{Polytropic star}
\label{sec_polytropic}
Giving more realistic equation of state, we present a neutron star solution in the bigravity theory.
We then discuss its mass and radius in order to give a constraint on the theory 
or the coupling constants by comparing them with observed values.

\begin{figure}[tbp]
\centering
\includegraphics[width=7cm,angle=0,clip]{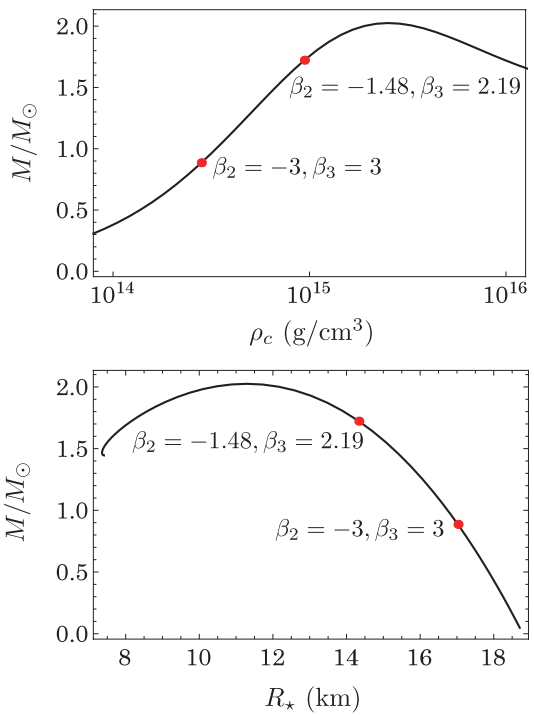}
\caption{
$\rho_c$-$M_\star$ and $R_\star$-$M_\star$ relations
for neutron stars with the polytropic equation of state \eqref{eos}.
The black solid lines are obtained in GR or in Class [II].
The maximum mass in Class [I], which is shown by the red dots
with $\beta_2=-3,\beta_3=3$ and $\beta_2=-1.48,\beta_3=2.19$,
depends on the coupling constants,
}
\label{fig_M_relations}
\end{figure}

\begin{figure}[tbp]
\centering
\includegraphics[width=7cm,angle=0,clip]{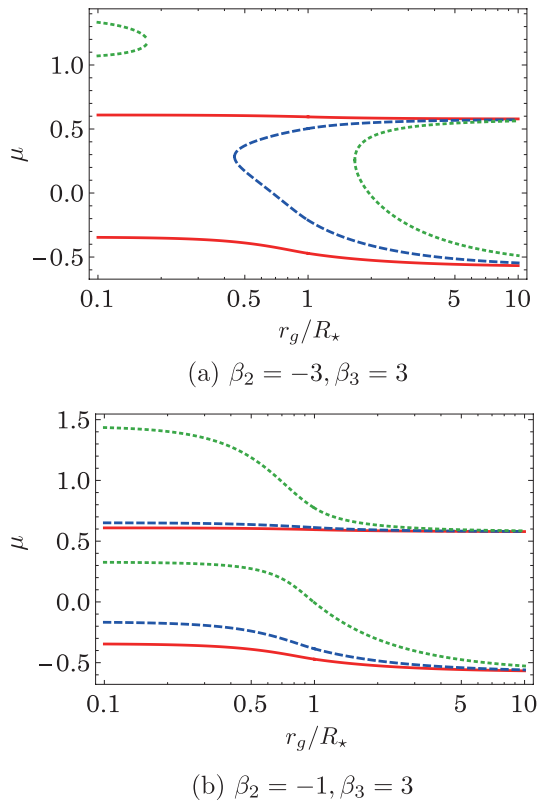}
\caption{
We set $m_g=m_f$, and (a) $\beta_2=-3,\beta_3=3$  (Class [I])
and (b) $\beta_2=-1,\beta_3=3$ (Class [II]).
We choose $\rho_c=1.71\times 10^{14}$ g/cm$^3$ (red solid curves), 
$3.35\times 10^{14}$ g/cm$^3$ (blue dashed curves)
and $18.9\times 10^{14}$ g/cm$^3$ (green dotted curves),
whose star masses are $0.6M_{\odot}$, $1.0M_{\odot}$ 
and $2.0M_{\odot}$, respectively.
}
\label{fig_mu_neutron_star}
\end{figure}

We assume  a simple polytropic-type equation of state
\begin{align}
P=\mathcal{K} \rho^2\,,
\label{eos}
\end{align}
where we set $\mathcal{K}=1.5\times 10^5$ [cgs].
In the massless limit of the graviton, we have two decoupled Einstein equations.
Then the $f$-metric is given by the Minkowski spacetime because there is no $f$-matter,
For $g$-spacetime, we have the same neutron star solution as that in GR.
We present $\rho_c$-$M_\star$ and $R_\star$-$M_\star$ relations in Fig.~\ref{fig_M_relations},
where $\rho_c=\rho_g(0)$ is the central density.
We find that the maximum mass of a neutron star is about 2$M_{\odot}$,
where $M_{\odot}$ is the solar mass, 
for the above equation of state.
This result is  obtained in GR but also it is the case for Class [II] in bigravity
because we always find the regular solution for $\mu(r)$ in the whole 
coordinate range ($0\leq r <\infty$).
We show some examples for the same coupling constants (\ref{coupling_class2}) 
with several values of the central density $\rho_c$ in Fig.~\ref{fig_mu_neutron_star}.

However, for Class [I], we find the additional constraint to find the regular $\mu(r)$
as we expect from the result in the previous subsection.
We also present some examples of $\mu(r)$ for the same coupling constants (\ref{coupling_class1}) 
with several values of  $\rho_c$ in Fig. \ref{fig_mu_neutron_star}.
This figure shows there is no regular solution of $\mu(r)$ in the whole region if the density 
$\rho_c$ is larger than $2.8\times 10^{14}$ g/cm$^3$.
This upper limit of the density does not reach the central density with the maximum mass of neutron star 
in GR (see Fig.~\ref{fig_M_relations}).
Hence this limit of  $\rho_c$
provides the maximum mass of a neutron star in Class [I],
which is much smaller than that in GR (or in Class [II]).
In Fig. \ref{fig_maximum_mass},
the maximum masses in Class [I] are shown by the contour lines.
The maximum mass is larger as the parameters $\{ \beta_2,\beta_3 \}$
come close to $\beta_2=-\sqrt{\beta_3}$.
The upper bound of the maximum mass in Class [I] is at most $1.72 M_{\odot}$,
which is realized at $\beta_2\simeq -1.48$ and $\beta_3 \simeq 2.19$.
Hence the maximum mass in Class [I] does not reach $2M_{\odot}$,
which may be inconsistent with  the existence of the $2M_{\odot}$ neutron star \cite{obs_neutron_star1,obs_neutron_star2,obs_neutron_star3,obs_neutron_star4} .
One might find a $2M_{\odot}$ neutron star in Class [I] if we modify 
the equation of state, but it will give a strong constraint on the coupling constants
in the theory. 

As for the compactness, we find 
\bea
\begin{array}{lll}
{\displaystyle {GM_\star\over R_\star}}\Big{|}_{\rm ub}=
&
0.18 &~~~{\rm for ~Class [I]}\,,
\\[1em]
{\displaystyle{GM_\star\over R_\star}}\Big{|}_{\rm max}=
&0.31 &~~~{\rm  for ~Class [II]}\,.
\\
\end{array}
\ena
Although both values are so far consistent with observations, 
the coupling constants in Class [I] may be restricted again because the above value is
just the upper bound.

\section{Regular compact objects : Numerical results}
\label{sec_relativistic_star}
In this section, we numerically solve the basic equations 
under the metric ansatz \eqref{g-metric} and \eqref{f-metric} with a $g$-matter field.
We find a relativistic star solution and confirm the previous results obtained in the massless limit
 when the graviton mass is sufficiently small. 

We numerically integrate Eqs.
\eqref{g-matter_con}, \eqref{diff_eq} and \eqref{Jacobian} outwards from the center
$r=0$.
The constraint equation \eqref{algebraic_eq} is used to evaluate 
the boundary values at the center.
Since it must be satisfied in the region of $r>0$ too,
we use this constraint to check the accuracy of our numerical solutions
in $r>0$.

Since the equations are seemingly singular at $r=0$,
We start our calculations from $r=0+\delta r$ with $\delta r\ll 1$.
All variables are expanded around $r=0$ as
\begin{align}
X=\sum_{n=0} \frac{1}{n!}X^{(n)}(0) r^n\,,
\label{variable_r0}
\end{align}
where $X^{(n)}(0)$ is the $n$-th derivative of the variable $X$ at $r=0$.

Here, by use of the freedom of  time coordinate rescaling,  
 we choose $N_g(0)=1$
without loss of generality.
\footnote{ 
Although it gives  $N_g(\infty)\neq 1$,
if we wish to find the boundary condition  $N_g(\infty)= 1$, 
we redefine new lapse functions as
\bea
\tilde N_g(r)={N_g(r)\over N_g(\infty)}
\,,~~
\tilde N_f(r)={N_f(r)\over N_g(\infty)}
\ena
and new time coordinate as 
\bea
 \tilde t = N_g(\infty) t 
 \ena
 New metrics defined by $\tilde N_g$, 
 $\tilde N_f$ and $\tilde t$ satisfy the boundary condition $\tilde{N}_g(\infty)=1$ at infinity.
}
We determine the values of variables at $r=\delta r$ 
by using up to second order of \eqref{variable_r0}.

In this section, we focus only on the branch A solution since we are interested in an asymptotically flat spacetime.
We will give some remarks for the branch B, which 
gives an asymptotically AdS spacetime,  in Appendix \ref{other_sol}.

\subsection{A uniform density star}

We first discuss a uniform density star, i.e., 
$\rho_g=$ constant. 
The dimensionless parameters characterizing the star are
\begin{align}
\kappa_g^2\rho_g/m_{\rm eff}^2\,,\quad
P_g^{(0)}(0)/\rho_g\,,
\label{star_para_g}
\end{align}
where we have defined 
\begin{align}
m_{\rm eff}^2=m_g^2+m_f^2\,,
\end{align}
which gives the effective graviton mass
on the homothetic spacetime.
The first parameter in (\ref{star_para_g}) is evaluated as 
\bea
{\kappa_g^2\rho_g\over m_{\rm eff}^2}={6GM_\star\over R_\star}\left({m_{\rm eff}^{-1}
\over R_\star}\right)^2
\,,
\ena
which is much larger than unity because  $m_{\rm eff}^{-1}$ is 
the Compton 
wavelength of the graviton and then it must be a cosmological scale.

Once the parameters \eqref{star_para_g} are given,
 the proper value of $\mu(0)$ is determined by a shooting method 
 to adjust the correct boundary condition 
\eqref{boundary_cond} at infinity 
as well as the asymptotic flatness.
Then all coefficients in Eq. \eqref{variable_r0} are fixed by 
this $\mu(0)$ from the expanded basic equations order by order,

We use $\mu_0$ as the center value of $\mu(0)$ in the case of massless limit.
When the value of the graviton mass is sufficiently small,
the proper value of $\mu(0)$ is close to $\mu_0$.
Hence, we start to search for $\mu(0)$ near $\mu_0$
to find a regular solution with the correct boundary condition.

To check the boundary conditions at infinity,
we evaluate the eigenvalues of $\gamma^{\mu}{}_{\nu}$, i.e., 
\begin{align}
\lambda_0&:=\frac{KN_f}{N_g}\,,\\
\lambda_1&:=\frac{Kr_f'/F_f}{1/F_g}\,,\\
\lambda_2&=\lambda_3:=\frac{Kr_f}{r}=K(1+\mu)\,.
\end{align}
If all eigenvalues approach the same constant $K$ as $r\rightarrow \infty$, 
the solution is asymptotically homothetic.
Then the  $\gamma$ energy-momentum tensor will become
a ``cosmological" constant ($\Lambda_g$) term at infinity.
We find our solution with an asymptotic flatness, 
if  $\Lambda_g=0$, which we have assumed for our coupling constants.

\subsubsection{\rm Class [I]}
As an example in Class [I], we choose the same coupling constants as before, i.e., 
\begin{align}
\Lambda_g&=0\,,\quad m_g=m_f\,, \quad
\beta_3=-3\,,\quad \beta_4=3\,.
\label{coupling_choice}
\end{align}

\begin{figure}[tbp]
\centering
\includegraphics[width=7cm,angle=0,clip]{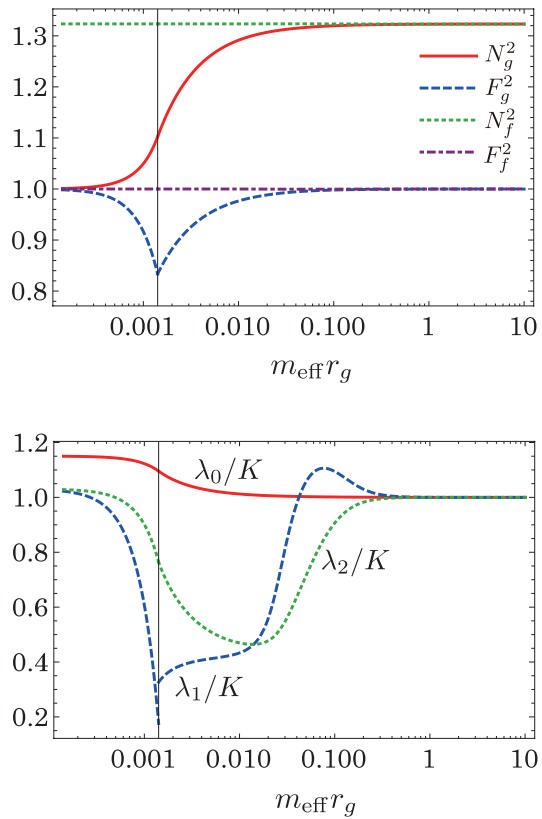}
\caption{
A typical solution for the branch A.
We set $P_g(0)/\rho_g=5 \times 10^{-2}$.
The shooting parameter is tuned to be $\mu(0)=0.03093$.
The vertical bar represents the star surface ($R_{\star}/m_{\rm eff}^{-1}=0.00141288$).
}
\label{fig_A}
\end{figure}

The branch A solution approaches an asymptotically flat 
 homothetic spacetime.
In Fig. \ref{fig_A}, we show a numerical solution by setting
$\kappa_g^2\rho_g/m_{\rm eff}^2=2.5\times 10^{5}$,
\footnote{This value is too small for a realistic neutron star with 
a massive graviton responsible for the present accelerating 
expansion of the Universe, for which we have $\kappa_g^2\rho_g/m_{\rm eff}^2\sim
 10^{43}$. However, because of the technical reason for numerical calculation,
  we choose the  above value. For the realistic value, we expect that the solution may be 
  closer to the case of the massless limit.}
for which the typical value of the Vainshtein radius 
is given by
\begin{align}
R_{\rm V}:=(GM_{\star}/m_{\rm eff}^2)^{1/3}
\sim 30 R_{\star}\,.
\end{align}
GR is recovered within the Vainshtein radius.

\begin{figure}[tbp]
\centering
\includegraphics[width=7cm,angle=0,clip]{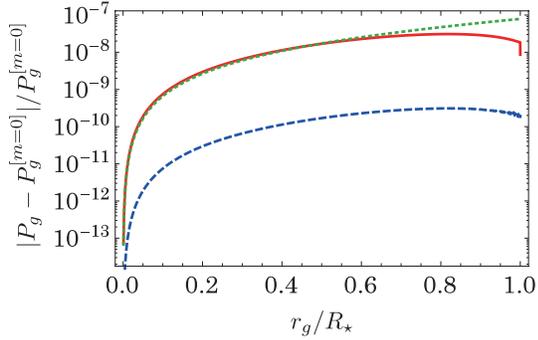}
\caption{
$|P_g-P_g^{[m=0]}|/P_g^{[m=0]}$
where $P_g$ is the numerical solution with a finite mass
and $P_g^{[m=0]}$ is the solution in massless limit.
We set $P_g(0)/\rho_g=5 \times 10^{-2}$
and \eqref{coupling_choice} with $\kappa_g^2\rho_g/m_{\rm eff}^2=2.5\times 10^{5}$
(the red solid curve),
 $\kappa_g^2\rho_g/m_{\rm eff}^2=2.5\times 10^{7}$
(the blue dashed curve)
and
\eqref{coupling_choice2} with $\kappa_g^2\rho_g/m_{\rm eff}^2=2.5\times 10^{5}$
(the green dotted curve).
We note $P_g-P_g^{[m=0]}>0$ for \eqref{coupling_choice},
while $P_g-P_g^{[m=0]}<0$ for \eqref{coupling_choice2}.}
\label{fig_difference}
\end{figure}

We note $\lambda_1$ is discontinuous at the star surface $R_{\star}$.
It is because the discontinuity of the matter distribution leads 
the discontinuity of $r_f'$ as seen in Eq. (\ref{Jacobian}).
This discontinuity disappears when we discuss a continuous matter distribution
such as a polytropic star \eqref{eos} as shown in Fig. \ref{fig_polytropic}.

Changing the central value of the pressure $P_g(0)/\rho_g$, we find 
the solution disappears for $P_g(0)/\rho_g>0.0665$.
It is consistent with the argument in the massless limit, in which 
the critical value is given by  $P_g(0)/\rho_g=1/15\approx 0.06667$.
Hence even in the case with a finite graviton mass, there exists
 a critical value of the pressure 
beyond which a regular star solution does not exist.

If we choose the larger value of the parameter as $\kappa_g^2\rho_g/m_{\rm eff}^2=2.5\times 10^{7}$,
the solution exists for $P_g(0)/\rho_g>0.0666$, which is closer to 
the value in the massless limit.
Hence, we expect that the massless limit approximation 
is valid for the realistic value $\kappa_g^2\rho_g/m_{\rm eff}^2\sim
 10^{43}$.

If the solution exists, 
the inner structure of star as well as the gravitational field
are restored to the result of GR because of the Vainshtein mechanism.
We find differences between our numerical solution and
the semi-analytic solution in massless limit are very small 
as shown one example of the pressure $P_g$ in Fig. \ref{fig_difference}.
This fact also confirms the validity of the massless limit approximation 
if the graviton mass is sufficiently small.
We conclude that the bigravity for Class [I] cannot reproduce the result in GR
beyond the critical value of $P_g(0)/\rho_g$.

\subsubsection{\rm Class [II]}
As an example in Class [II], we choose one of the previous coupling constants, i.e., 
\begin{align}
\Lambda_g=0\,, \quad m_g=m_f\,, \quad \beta_3=1\,, \quad \beta_3=3
\label{coupling_choice2}
\end{align}
and we set
\begin{align}
\kappa_g^2\rho_g/m_{\rm eff}^2=2.5\times 10^5\,.
\end{align}
In this case, we can find a regular star for any values of $P_g(0)$.
The solution is almost the same as the massless limit (or GR) 
as shown in Fig. \ref{fig_difference}.
We conclude that in the bigravity theory in Class [II] 
the results in  GR are recovered and 
the Vainshtein mechanism holds even in a strong gravity limit.

\subsection{ Polytropic star}
\begin{figure}[h]
\centering
\includegraphics[width=7cm,angle=0,clip]{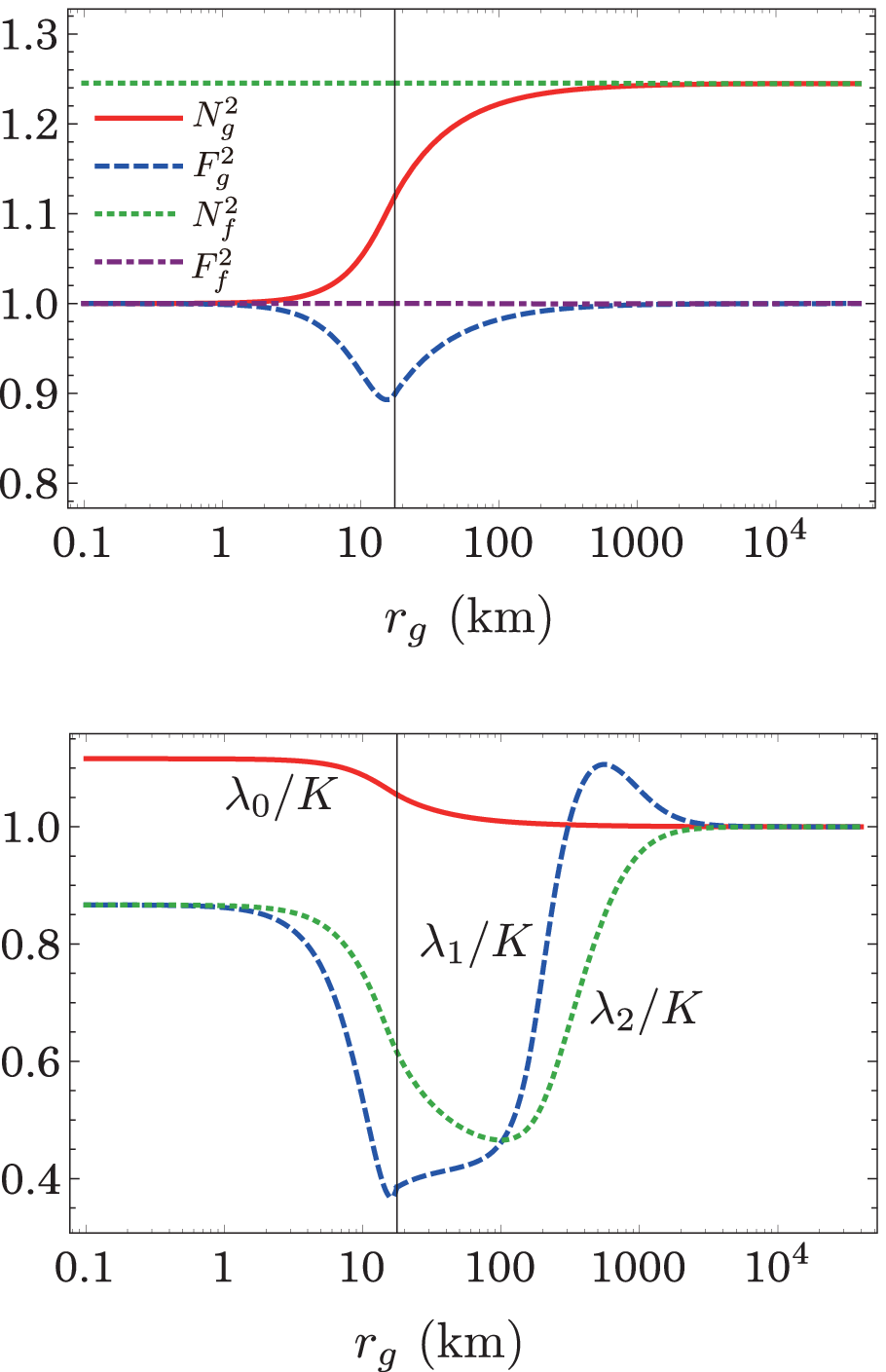}
\caption{
We set $\rho_c=1.71 \times 10^{14}\; {\rm g/cm}^3$
and $m_{\rm eff}^{-1}=10^4$ km,
for which the mass of the neutron star is $0.601 M_{\odot}$.
The shooting parameter is tuned to be $\mu(0)=-0.13334$.
The vertical bar represents the star surface ($R_{\star}=17.7$ km).
}
\label{fig_polytropic}
\end{figure}

For a neutron star with a realistic equation of state,
we can also confirm the above results, i.e. the massless limit is 
valid.
Here we again assume
the polytropic equation of state \eqref{eos}.

One typical example of the solutions
in Class [I] is shown in Fig. \ref{fig_polytropic},
where we choose the coupling constants as \eqref{coupling_choice} and
\begin{align}
\rho_c=1.71\times 10^{14}\;{\rm g/cm}^3\,,\quad
m_{\rm eff}^{-1}=10^4\;{\rm km}\,,
\end{align}
We find a neutron star solution with
\begin{align}
M_{\star}=0.601 M_{\odot}\,,\quad
R_{\star}=17.7\; {\rm km}\,,
\end{align}
which is the same as those in the massless limit.
Our numerical calculation shows that  increasing the central density $\rho_c$,
the solution exists only 
for $M_{\star}\lsim  0.882 M_{\odot}$
for  the coupling constants \eqref{coupling_choice}.
We have obtained $M_{\star} \lesssim   0.886 M_{\odot}$ in the massless limit.
If we choose the larger value of the Compton wave length of the graviton as
$m_{\rm eff}^{-1}=10^5$ km,
the mass upper limit increases as $M_{\star}\lesssim  0.884 M_{\odot}$,
which is closer to the value in the massless limit.

For Class [II], we always find the same solution as that in GR.
As a result,  as the case of a uniform-density star,
we confirm that
the massless limit solution 
is a good approximation
for the sufficiently small graviton mass.

~~
\vskip 1cm

\section{Concluding remarks}
\label{summary}
Assuming static and spherically symmetric spacetimes, 
We have presented a relativistic star solution in the bigravity theory.
For simplicity, we have considered only $g$-matter fluid
and given  only asymptotically flat solutions in the text.
Some solutions with the other conditions are discussed in Appendix \ref{other_sol}.

First we obtain the solutions under the massless limit approximation in Sec. \ref{sec_massless_limit}.
Then, by solving the basic equations numerically without 
the approximation  in Sec. \ref{sec_relativistic_star},  we confirm such an approximation is valid 
since the graviton mass, if it exists,  must be sufficiently small.

We find that 
the coupling constants are classified into two classes: Class [I] and Class [II].
For both classes, the Vainshtein screening is found in the weak gravitational field.
However, when we take into account a relativistic effect,
the Vainshtein screening  mechanism may not work in some strong gravity regime
in Class [I].
In fact, to find a regular function of $\mu(r)$ in Class [I], 
the central pressure is constrained, 
and as a result, the maximum mass 
is much smaller than that in GR
as shown in Fig. \ref{fig_maximum_mass}.
Beyond this maximum mass, 
the Vainshtein mechanism does not work well
since GR solution is not obtained.

On the other hand, there is no additional constraint for Class [II],
and the structure of star as well as the gravitational field 
are restored to those in GR for the expected small graviton mass.
The Vainshtein screening mechanism works well in Class [II].

In Table \ref{table_summary},
we summarize our results.

\begin{widetext}

\begin{table}[H]
\begin{center}
  \begin{tabular}{|c|c|c|c|c||}
\hline 
class &\multicolumn{2}{|c|}{Class [I]}&\multicolumn{2}{|c|}{Class [II]}\\ 
\hline 
\hline 
coupling &\multicolumn{4}{|c|}{ $\beta_3>1$ \& $d_1 +d_2 \, \beta_2<0$}\\
\cline{2-5}
constants&\multicolumn{2}{|c|}{$\beta_2<-\sqrt{\beta_3}$ }&
\multicolumn{2}{|c|}{$-\sqrt{\beta_3}<\beta_2\lsim \sqrt{\beta_3}$ }\\
\hline
equation of state&uniform density&polytrope&uniform density&polytrope
\\
\hline
mass&-&$M_{\rm ub}= 1.72 M_\odot$&-&$M_{\rm max}=2.03M_\odot$
\\
\hline 
&&&&
\\[-1em]
compactness
&$\displaystyle{{M_\star\over R_\star}}\Big{|}_{\rm ub}$=0.23
 & $\displaystyle{{M_\star\over R_\star}}\Big{|}_{\rm ub}$ =0.18
 & $\displaystyle{{M_\star\over R_\star}}\Big{|}_{\rm max}$= 0.44
&$\displaystyle{{M_\star\over R_\star}}\Big{|}_{\rm max}$= 0.31
\\
\hline \hline 
 \end{tabular}
    \caption{
The maximum masses and the maximum compactnesses
for Class [I] and Class [II].
For Class [I], the maximum values depend on the coupling constants. 
Then we only show the upper bounds
which are realized for $\beta_2\simeq -1.48,\beta_3\simeq 2.19$.}
\label{table_summary}
\end{center}
\end{table}
\end{widetext}

The result suggests that
Class [II] is favored from the existence condition of a neutron star.
As the necessary condition of Class [II], 
the parameters should satisfy
\begin{align}
\beta_2^2-\beta_3\leq 0\,,
\label{astrophysical_constraint}
\end{align}
as shown in Fig. \ref{fig_maximum_mass}.
However, 
those parameters should happen to satisfy
\begin{align}
\beta_2^2-\beta_3>0\,,
\end{align}
from the cosmological point of view,
which constraint comes from to find 
a stable solution in the early Universe in bigravity \cite{KA_KM_RN}.
There is no intersection of the parameters 
because the boundaries of Class [II] and of the cosmological constraint 
coincide exactly.
If we take the parameters in  Class [I] from the cosmological constraint,
the equation of state of the star will be strongly constrained
to find a two solar mass neutron star.
Conversely, if we assume Class [II] from the astrophysical point of view,
the problem of ghost or gradient instability may reappear in the early Universe. 

There is another problem in Class [II] parameters.
Since we have started to discuss the bigravity theory in order to explain 
the present acceleration of the Universe, 
the parameters (or coupling constants) should predict 
the existence of a positive effective cosmological constant ($\Lambda_g>0$).
If we impose the same conditions on the coupling constants as discussed in \cite{with_twin_matter},
$\{b_i\}$'s are given by two coupling constants $c_3$ and $c_4$.
The existence condition of de Sitter solution as well as Minkowski solution yields 
\bea
2c_3^2+3c_4>0
\,,
\ena
which excludes the possibility of (\ref{astrophysical_constraint}).
Hence,
if we assume the Minkowski spacetime is a vacuum solution,
Class [II] cannot admit the de Sitter solution as another vacuum solution as well,
thus the acceleration of the Universe cannot be explained in the bigravity.

In this paper, we have assumed that 
both static $g$- and $f$-spacetimes are static with respect to the same time coordinate $t$,
and the St\"uckelberg field $\mu$ is also  static.
However there is a possibility such that
the existence of the critical value in Class [I] might be caused 
by the above simple ansatz.
The static ansatz of the St\"uckelberg field
may not be necessary to obtain an (approximate) static spacetime.
In fact, in the case of cosmology, 
a homogeneous configuration of the St\"uckelberg field
leads an instability,
while the inclusion of an inhomogeneity in the St\"uckelberg field
gives a stable solution,
 which describes an (approximate) homogeneous spacetime due to the 
 Vainshtein screening \cite{KA_KM_RN}.
Hence, to draw a final conclude about the existence of a massive neutron star 
(and also a black hole solution),
relaxing
the static ansatz of the St\"uckelberg field,
we should extend our analysis to the spacetime with dynamical St\"uckelberg fields,
which we leave for our future work.

\section*{Acknowledgments}
We would like to thank Kotaro Fujisawa, Takashi Nakamura, and Mikhail S. Volkov 
for useful discussions and comments.
 This work was supported in part by Grants-in-Aid from the 
Scientific Research Fund of the Japan Society for the Promotion of Science 
(No. 25400276 and No. 15J05540).


\appendix


\section{Weak gravity approximation}
\label{sec_weak_gravity}
In this appendix, 
we discuss the constraint on the coupling constants
in order to find a successful Vainshtein screening mechanism
in a weak gravity system.
We assume only $g$-matter field for simplicity, 
and chose the gauge $r_g=r$.
We define new variables by
\begin{align}
N_g&=e^{\Phi_g}\,,\quad
F_g=e^{-\Psi_g}\,, \\
N_f&=e^{\Phi_f}\,,\quad
F_f=e^{-\Psi_f}\,,
\end{align} 
with the following conditions:
\begin{align}
|\Phi_g|,|\Psi_g|,|\Phi_f|,|\Psi_f| &\ll 1\,,
\label{weak_gravity1}
\\
|r \Phi_g'|,|r \Psi_g'|,|r \Phi_f'|,|r \Psi_f'| &\ll 1\,,
\label{weak_gravity2}
\end{align}
where 
a prime denotes the derivative with respect to $r$.
From the basic equations, we find a septic equation for $\mu$ as
\begin{align}
&\mathcal{C}_{m^2}(\mu)+\mathcal{C}_{\Lambda}(\mu) 
+\mathcal{C}_{\rm matter}(\mu)=0 \label{mu_eq2}\,,
\end{align}
where $\mathcal{C}_{m^2},\mathcal{C}_{\Lambda}$  and $\mathcal{C}_{\rm matter}$ 
are explicitly defined in \cite{KA_KM_RN}.
These terms have typical magnitudes given by 
\begin{align*}
\mathcal{C}_{m^2}&\sim m_{\rm eff}^2 \times \mathcal{O}(\mu^7)
\,,\\
\mathcal{C}_{\Lambda}&\sim \Lambda_g \times \mathcal{O}(\mu^5)
\,,
\end{align*}
and the last term is given by
\begin{align}
\mathcal{C}_{\rm matter}=\frac{6GM_{\star}}{r^3}(1+\mu)^2(1-\beta_3\mu^2)
\,,
\label{C_matter}
\end{align}
where $M_{\star}$ is the gravitational masses of the $g$-matter.

There is a root of Eq.~\eqref{mu_eq2} with $\mu \rightarrow 0$ as $r\rightarrow \infty$,
which is the asymptotically homothetic branch.
Such a branch should be extended inward without any singularity.
As discussed in \cite{Babichev,Enander},
the branch with $\mu = 0$ at $r = \infty$
reaches to $\mu \rightarrow - 1/\sqrt{\beta_3}$ in the range of  $r\ll R_V$,
where we find a successful Vainshtein screening.

Although
we cannot find analytic roots $\mu(r)$ of the septic equation \eqref{mu_eq2}, 
we can easily find a inverse function $r(\mu)$
because $r$ appears only in $\mathcal{C}_{\rm matter}$ as the form \eqref{C_matter}.
The result indicates that the function $r(\mu)$ is a single-valued function.
However, the function $\mu(r)$ is not a single-valued function,
if there is an extremal value of the function $r(\mu)$, i.e., $dr/d\mu=0$.
The point of $dr/d\mu=0$ corresponds to a curvature singularity.
Hence a regular solution must be given by a monotonic function $\mu(r)$
in the domain $R_I<r<\infty$,
where $R_I$ is a typical length, if it exists,
 below which the weak gravity approximation is not valid.

As discussed in \cite{Enander},
we find the parameter constraint as follows:
Since the function $\mu(r)$ should be monotonic,
the function is approximated by
\begin{align}
\mu=-1/\sqrt{\beta_3}+\delta \mu(r)\,,
\end{align}
with $1\gg \delta \mu>0$ in $r\ll R_V$.
Substituting this expression into \eqref{mu_eq2},
we find
\begin{align}
&\mathcal{C}_{m^2}|_{\mu=-1/\sqrt{\beta_3}}
+\mathcal{C}_{\Lambda}|_{\mu=-1/\sqrt{\beta_3}}
\nn
&\approx -\frac{12GM_{\star}}{r^3}
( 1-1/\sqrt{\beta_3} )^2 \sqrt{\beta_3}\delta \mu
\,.
\end{align}
Since the right hand side is negative, the necessary condition is given by
\begin{align}
&-\left(\mathcal{C}_{m^2}|_{\mu=-1/\sqrt{\beta_3}}
+\mathcal{C}_{\Lambda}|_{\mu=-1/\sqrt{\beta_3}}\right)
\nn
&=\frac{2}{\beta_3^{5/2}}(\beta_2-\sqrt{\beta_3})
(d_1+\beta_2 d_2)>0\,,
\label{Vainshtein_constraint}
\end{align}
where
\begin{align}
d_1&:=-6m_g^2\sqrt{\beta_3}(1-\sqrt{\beta_3})^2
\nn
&\quad \;\;
+m_f^2(1-6\sqrt{\beta_3}+13\beta_3-6\beta_3^{3/2})
\label{def_d1}
\\
&\quad \;\;
+(1-\sqrt{\beta_3})^2(-1+4\sqrt{\beta_3})\Lambda_g\,,
\\
d_2&:=
3m_g^2(1-\sqrt{\beta_3})^2+m_f^2(1-6\sqrt{\beta_3}+3\beta_3)\,.
\label{def_d2}
\end{align}

However the constraint \eqref{Vainshtein_constraint}
is not sufficient, 
because it does not guarantee that
the function $\mu(r)$ is a single-valued function in the domain $R_I<r<\infty$,
which is guaranteed by $r(\mu)$ has no extremal value in $-1/\sqrt{\beta_3}<\mu<0$.
We must impose 
$dr(\mu)/d\mu>0$ for any $\mu$ with $-1/\sqrt{\beta_3}<\mu<0$
which gives further constraint on the coupling constants.

\begin{figure}[tbp]
\centering
\includegraphics[width=8cm,angle=0,clip]{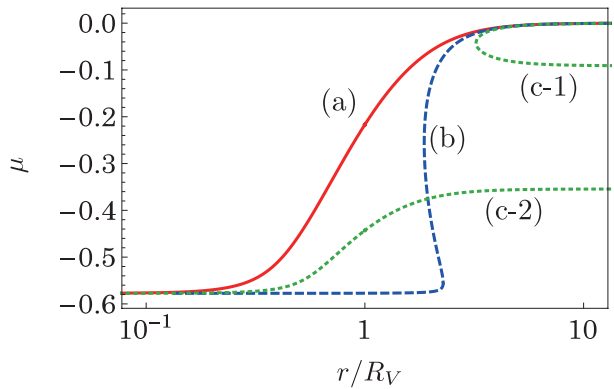}
\caption{Examples of the root of \eqref{mu_eq2}.
We set $m_g=m_f,\Lambda_g=0$ and
(a) $\beta_2=-3,\beta_3=3$ (red solid curve), 
(b) $\beta_2=1.73,\beta_3=3$ (blue dashed curve), 
and 
(c) $\beta_2=7,\beta_3=3$ (green dotted curves).
Only the case (a) gives a regular asymptotically flat solution.
}
\label{fig_mu_examples}
\end{figure}

\begin{figure}[tbp]
\centering
\includegraphics[width=8cm,angle=0,clip]{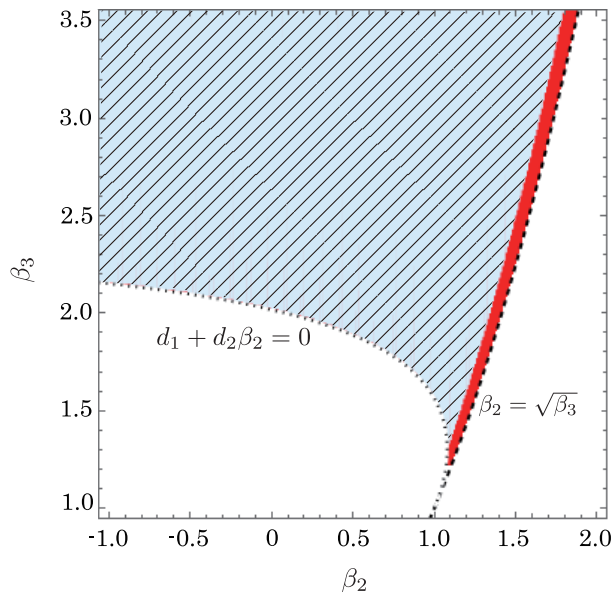}
\caption{The parameter space for a successful Vainshtein screening.
We set $m_g=m_f, \Lambda_g=0$.
The colored (the light-blue and red) regions  satisfy $\beta_2-\sqrt{\beta_3}<0,\, 
d_1+\beta_2d_2<0$.
However, only the hatched light-blue region satisfies the condition such that 
$\mu(r)$ is a single-valued function.
}
\label{fig_Vainshtein_constraint}
\end{figure}

Three examples of the solution $\mu(r)$ are shown in Fig.~\ref{fig_mu_examples}:
(a) $\beta_2=-3,\beta_3=3$,
(b) $\beta_2=1.73,\beta_3=3$,
and
(c) $\beta_2=7,\beta_3=3$.
The case (a) and (b) satisfy
\begin{align}
\beta_2-\sqrt{\beta_3}< 0\,,
\quad 
d_1+\beta_2d_2< 0\,,
\end{align}
while the case (c) satisfies
\begin{align}
\beta_2-\sqrt{\beta_3}> 0\,,
\quad 
d_1+\beta_2d_2> 0\,.
\end{align}
For both (a) and (b), 
the branch of $\mu\simeq -1/\sqrt{\beta_3}$ in $r\ll R_V$ connects
the branch of $\mu=0$ at $r=\infty$.
However, the case (a) gives the single-valued function $\mu(r)$,
while the case (b) is not.
It indicates that the ratio of two radial coordinates are not single-valued function
\footnote{
A similar behaviour is found in the context of cosmology,
for which the ratio of cosmic times is not single-valued function
\cite{with_twin_matter}.}.
For the case (c),
there are two curves (c-1) and (c-2)
and these are disconnected.
Note that, the branch (c-2)
can be extended to infinity.
This branch is not an asymptotically Minkowski solution,
but an asymptotically AdS solution
similarly to the branch C
which will be discussed in Appendix \ref{sec_f_star}.

As a result, the parameter constraint is approximately given by
\begin{align}
\beta_2-\sqrt{\beta_3}\lesssim 0\,,
\quad 
d_1+\beta_2d_2< 0\,,
\end{align}
as shown in Fig.~\ref{fig_Vainshtein_constraint}.
The hatched light-blue region gives 
a successful Vainshtein screening solution.
We can show numerically that
there is no regular asymptotically homothetic solution 
in the narrow region along $\beta_2=\sqrt{\beta_3}$ (the red region),
in which $\mu(r)$ is not a single-valued function such as (b) in Fig.~\ref{fig_mu_examples},
and should then be excluded.


\section{Asymptotically non-flat solutions}
\label{other_sol}
In this Appendix, we analyze
asymptotically non-flat solutions.
We consider only the case such that 
the $\gamma$ energy-momentum tensor approaches to 
a cosmological constant at infinity.
There are two types of non-asymptotically flat spacetimes:
One is an asymptotically homothetic spacetime, and the other is 
an asymptotically non-homothetic one. 
We find  de Sitter, Minkowski or 
anti-de Sitter spacetime at infinity.
Here we have also assumed that a regular solution exists in the whole range of 
the radial coordinate $r$ ($0\leq r<\infty$),
which excludes a spacetime approaching the Nariai solution asymptotically.

When there exists a positive cosmological constant,
the cosmological  horizon may appear at $r\simeq \sqrt{3/\Lambda_g}$.
From the regularity condition at the horizon,
the metric variables should satisfy
\begin{align}
N_g=F_g=N_f=F_f=0\,,
\label{boundary_horizon}
\end{align}
at the horizon.
We regard the solution satisfies this boundary condition 
when we obtain $N_g^2,N_f^2,F_g^2,F_f^2 <10^{-4}$.

\subsection{Effective cosmological constants}
\label{appendix}

First, we summarize when the $\gamma$ energy-momentum tensor is reduced to
just a cosmological constant at infinity.
For the ansatz \eqref{g-metric} and \eqref{f-metric}, we find the eigenvalues 
$\{\lambda_0,\lambda_1,\lambda_2,\lambda_3\}$ of $\gamma^\mu_{~\nu}$ as 
\bea
\lambda_0 &:=&\gamma^t{}_t={KN_f\over N_g}\nonumber \\
\lambda_1 &:=&\gamma^r{}_r={Kr_f' F_g\over r_g' F_f}\nonumber \\
\lambda_2 &:=&\gamma^{\theta}{}_{\theta}
={Kr_f\over r_g}=K(\mu+1)\nonumber \\
\lambda_3 &:=&\gamma^{\varphi}{}_{\varphi}={Kr_f\over r_g}=K(\mu+1)
\ena
Then, the $\gamma$ energy-momentum tensor is given by
\begin{widetext}
\begin{align}
\frac{\kappa^2}{m^2}T^{[\gamma]t}_g{}_t&=-(b_0+2b_1\lambda_2+b_2\lambda_2^2)-\lambda_1(b_1+2b_2\lambda_2+b_3\lambda_2^2)\,,\\
\frac{\kappa^2}{m^2}T^{[\gamma]r}_g{}_r&=-(b_0+2b_1\lambda_2+b_2\lambda_2^2)-\lambda_0(b_1+2b_2\lambda_2+b_3\lambda_2^2)\,,
\\
\frac{\kappa^2}{m^2}
\left(T^{[\gamma] \theta}_g{}_{\theta}-T^{[\gamma] r}_g{}_{r}\right)&=
(\lambda_2-\lambda_0)(b_1+b_2(\lambda_1+\lambda_2)+b_3\lambda_1\lambda_2)\nn
&=(\lambda_2-\lambda_1)(b_1+b_2(\lambda_0+\lambda_2)+b_3\lambda_0\lambda_2)+(\lambda_1-\lambda_0)(b_1+2b_2\lambda_2+b_3\lambda_2^2)\,,
\\
T^{[\gamma] \theta}_g{}_{\theta}
&=
T^{[\gamma] \varphi}_g{}_{\varphi}
\end{align}
\end{widetext}
We then find in following three cases 
that the $\gamma$ energy-momentum tensor turns to be a cosmological constant:
\\
Case (i) 
\begin{align}
\lambda_0=\lambda_1=\lambda_2={\rm constant}\,,
\end{align}
Case (ii)
\begin{align}
\lambda_0=\lambda_2\,, \quad b_1+2b_2\lambda_2+b_3\lambda_2^2=0\,,
\end{align}
Case (iii)
\begin{align}
\lambda_1=\lambda_2\,, \quad b_1+2b_2\lambda_2+b_3\lambda_2^2=0\,.
\end{align}

We note that the equation $b_1+2b_2\lambda_2+b_3\lambda_2^2=0$ is equivalent to
\begin{align}
1+2\beta_2 \mu +\beta_3 \mu^2=0\,,
\end{align}
where we use $\lambda_2=K(1+\mu)$.

Case (i) gives an asymptotic homothetic spacetime,
i.e., an asymptotic de Sitter or anti-de Sitter spacetimes as well as 
an asymptotic Minkowski spacetime.
In addition, as we will show in the next subsection,
we also find a solution with a cosmological constant given by Case (ii).

\subsection{Relativistic star with $g$-matter}

Just for simplicity, we discuss a uniform-density star only with  $g$-matter fluid.
We use the parameters \eqref{coupling_class1} as an example for Class [I],
and parameters \eqref{coupling_class2} for Class [II].
We then choose 
\begin{align}
\kappa_g^2\rho_g/m_{\rm eff}^2=2.5 \times 10^5 \,,\quad
P_g(0)/\rho_g=5 \times 10^{-2}\,.
\label{pressure}
\end{align}
Since the central pressure (\ref{pressure}) is lower than the critical value,
we find a regular star solution both in Class [I] and in Class [II].
As discussed in Sec. \ref{sec_massless_limit},
there are two branches A and B.

\subsubsection{{\rm Branch A  (homothetic spacetime at infinity)}}
In the text, we consider the branch A without a cosmological constant,
in which  case,
the branch A solution approaches the Minkowski homothetic spacetime.
Here, we discuss asymptotic structures of branch A
when we introduce a non-zero cosmological constant.

\begin{figure}[tbp]
\centering
\includegraphics[width=7cm,angle=0,clip]{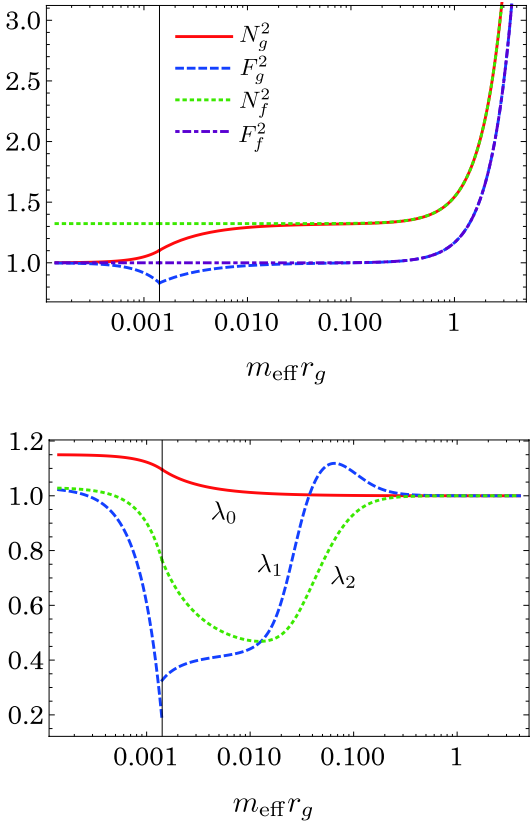}
\caption{The typical solution with a negative cosmological constant
for branch A.
We set $\Lambda_g=-m_{\rm eff}^2$
and $m_g^2=m_f^2,\beta_2=-3,\beta_3=3$.
The shooting parameter is tuned to be $\mu(0)=0.0305$.
The vertical var represents the surface of the star.
}
\label{fig_homothetic_AdS}
\end{figure}

\begin{figure}[tbp]
\centering
\includegraphics[width=7cm,angle=0,clip]{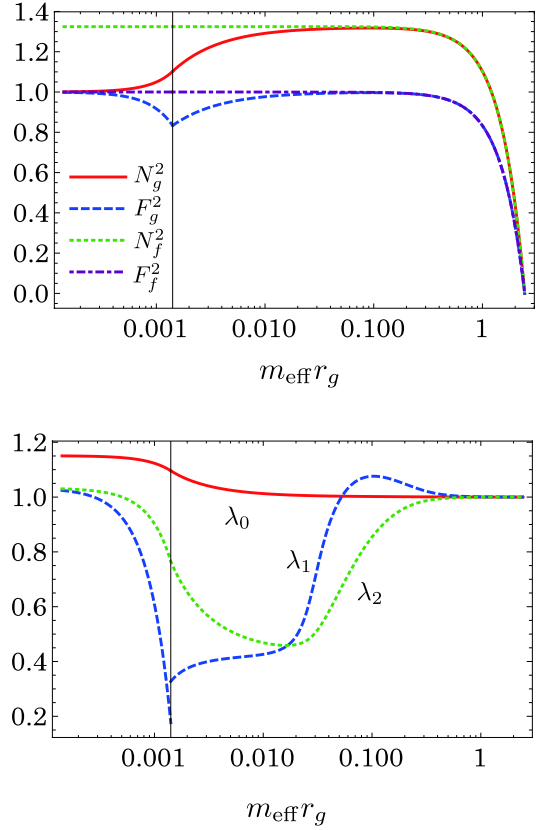}
\caption{The same figure as Fig. \ref{fig_homothetic_AdS}
in the case of a positive cosmological constant $(\Lambda_g=m_{\rm eff}^2)$.
The shooting parameter is tuned to be 
$\mu(0)=0.0326$
}
\label{fig_homothetic_dS}
\end{figure}

For the branch A, the results are the same both in Class [I] and in Class [II].
When we introduce a negative cosmological constant,
the solution approaches the homothetic anti-de Sitter spacetime 
at infinity as shown in Fig. \ref{fig_homothetic_AdS}
($N_g/N_f,F_g/F_f,r_g/r_f\rightarrow 1$).
For a positive cosmological constant,
when $2\Lambda_g \lesssim  3m_{\rm eff}^2$
(the Higuchi bound) is satisfied,
the solution seems to approach a homothetic de Sitter spacetime.
Since we cannot solve the basic equations beyond the cosmological horizon,
we cannot conclude definitely that the solution is asymptotically homothetic, but
 as shown in Fig. \ref{fig_homothetic_dS},
the solution seems to approach a homothetic spacetime 
because the eigenvalues coincide around $r\approx m_{\rm eff}^{-1}$ before the horizon.
However, if $2\Lambda_g   \gtrsim 3m_{\rm eff}^2$,
a regular solution disappears as discussed in the appendix of \cite{KA_KM_RN}.
As a result, the branch A always approaches a homothetic spacetime 
if the cosmological constant satisfies  $2\Lambda_g \lesssim  3m_{\rm eff}^2$.

\subsubsection{{\rm Branch B}}

\begin{figure}[tbp]
\centering
\includegraphics[width=7cm,angle=0,clip]{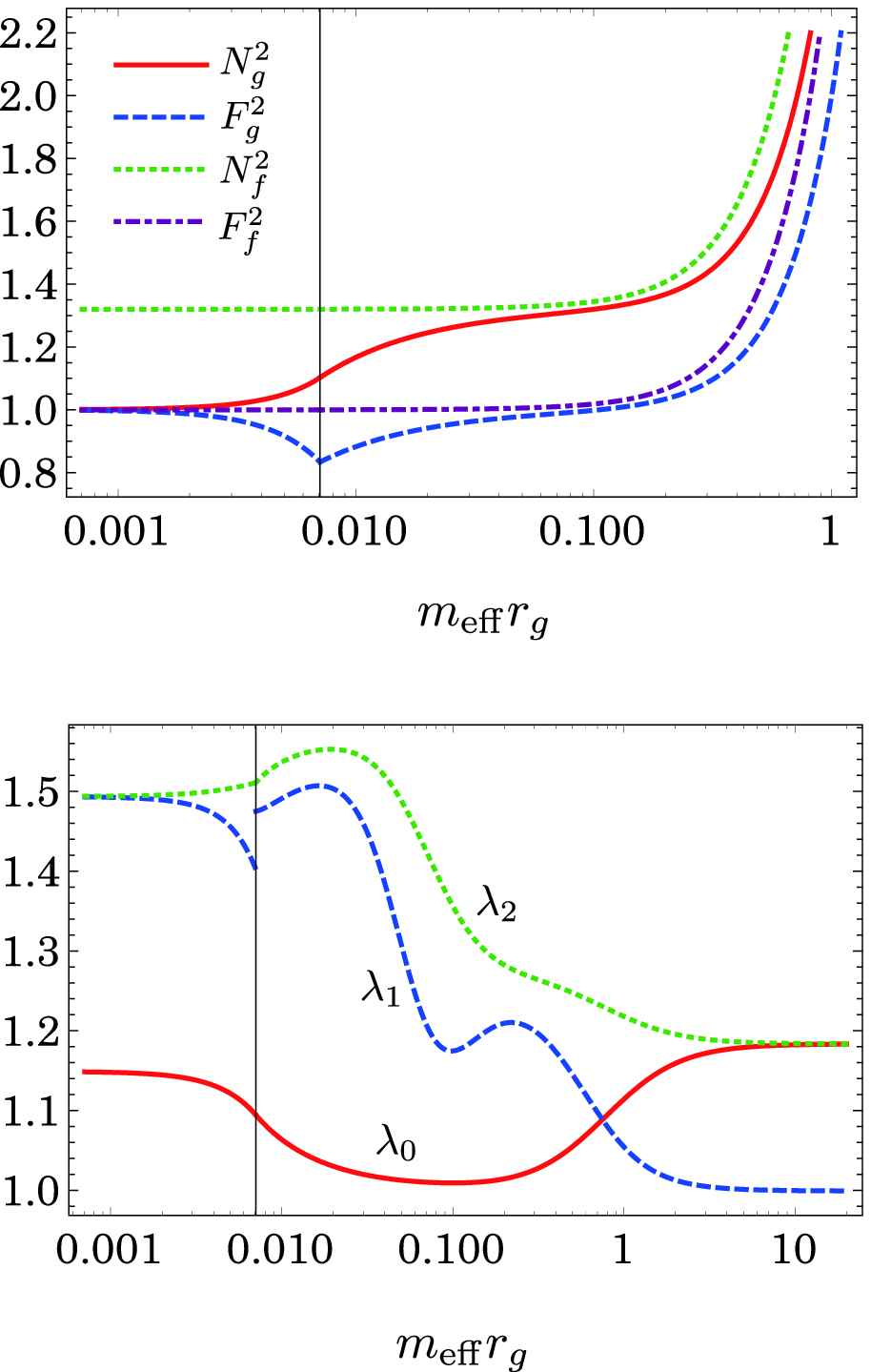}
\caption{The typical solution with a negative cosmological constant
for branch B in Class [I].
We set $m_g^2=m_f^2,\beta_2=-3,\beta_3=3$ and 
$\Lambda_g=-75 m_{\rm eff}^2\quad (m_{\rm eff}^{-1}=5\, \ell_{\rm AdS})$.
The shooting parameter is tuned to be $\mu(0)=0.4936$.
}
\label{fig_negative_cc}
\end{figure}

For the branch B, there is no regular solution in Class [II] for any cosmological constant.
On the other hand, in Class [I],
there exists a regular solution only if we introduce a negative cosmological constant
with $\ell_{\rm AdS}\lsim m_{\rm eff}^{-1}$ as shown in Fig. \ref{fig_negative_cc},
where $\ell_{\rm AdS}:=\sqrt{-3/\Lambda_g}$ is the AdS curvature radius.
Note that this solution is not asymptotically homothetic.
The eigenvalues $\lambda_0$ and $\lambda_2$ approach the same value
with satisfying $1+2\beta_2 \mu +\beta_3 \mu^2=0$,
for which the interaction term becomes just a cosmological constant 
as discussed in Appendix \ref{appendix}.
Although the $g$- and $f$-spacetimes are not homothetic at infinity,
both spacetimes approach asymptotically to some AdS spacetimes.

\subsection{Relativistic star with $f$-matter}
\label{sec_f_star}

Here, we discuss the effect of the $f$-matter field.
For simplicity, we assume $\rho_f \gg \rho_g$,
for which  we regard that the $g$-spacetime is almost vacuum.

The action of the bigravity is symmetric for $g$- and $f$-spacetimes under
the transformation
\begin{align}
g \leftrightarrow f\,, \quad
b_0\leftrightarrow b_4\,, \quad b_1 \leftrightarrow b_3\,.
\label{g_leftright_f}
\end{align}
Then the case only with $f$-matter
is equivalent to the case only with $g$-matter for
corresponding coupling constants under the transformation \eqref{g_leftright_f},
i.e., 
\begin{align}
\beta_2\rightarrow 1-\beta_2\,,\quad
\beta_3 \rightarrow 1-2\beta_2+\beta_3\,.
\end{align}
One can see that parameters in Class [II]
is still in Class [II] after the transformation \eqref{g_leftright_f}.
Therefore, the case only with $f$-matter in Class [II]
is equivalent to the case only with $g$-matter in Class [II],
which have already discussed in previous subsection.
Only non-trivial effect of $f$-matter exists in Class [I],
which we discuss here.

We briefly comment on the case of $\rho_g\sim \rho_f$.
For this case, the asymptotically homothetic branch cannot be extended inward
similarly to the solution (c-1) in Fig. \ref{fig_mu_examples} \cite{KA_KM_RN}.
Although the result presented in \cite{KA_KM_RN}
is only the case of Class [I],
we find the same behaviour even for Class [II].
One exceptional case is a homothetic solution.
If $\kappa_g^2T^{[{\rm m}]\mu}{}_{\nu}=K^2\kappa_f^2 \mathcal{T}^{[{\rm m}]\mu}{}_{\nu}$,
there exists a homothetic solution, i.e., $N_g=N_f,F_g=F_f$ and $\mu=0$,
for which the solution is identical to that in GR in the whole space region.

\subsubsection{Massless limit approximation}
In the massless limit,
the interior solution is given by
\begin{align}
F_g&=1\,, 
\\
N_g&=N_g(0)\,, 
\\
F_f&=\left(1-\frac{2\mathcal{GM}_{\star}}{\mathcal{R}_{\star}^3}r_f^2\right)^{1/2}\,, 
\\
N_f&=N_f(0)\frac{3F_f(\mathcal{R}_{\star})-F_f(r_f)}{3F_f(\mathcal{R}_{\star})-1}\,, 
\\
\frac{P_f(r_f)}{\rho_f(0)}&=
\frac{F_f(r_f)-F_f(\mathcal{R}_{\star})}{3F_f(\mathcal{R}_{\star})-F_f(r_f)}\,,
\end{align}
where we assume a uniform density for $f$ matter fluid.
The $g$-spacetime is just a Minkowski solution.
The exterior solution is given by
\begin{align}
F_g&=1 \,, \\
N_g&=N_g(0)\,,  
\label{massless_vac1}\\
F_f&=\left(1-\frac{2\mathcal{GM}_{\star}}{r_f} \right)^{1/2} \,, \\
N_g&=\frac{2N_f(0)}{3F_f(\mathcal{R}_{\star})-1}F_f(r_f)\,.
\label{massless_vac2}
\end{align}
where we define the gravitational mass by
\begin{align}
\mathcal{M}_{\star}=\int_0^{\mathcal{R}_{\star}} 4 \pi r_f^2 \rho_f dr_f \,.
\end{align}
and $\mathcal{R}_{\star}$ is the radius of the $f$-star measured in $f$-spacetime.
Similarly to the argument in Sec. \ref{sec_massless_limit},
the ratio must be
\begin{align}
\frac{N_f(0)}{N_g(0)}=\frac{2}{3F_f(\mathcal{R}_{\star})-1}\,.
\end{align}
The center value of $\mu$ is given by a root of
\begin{align}
&\quad\,
(3P_f(0)(2\beta_2-3\beta_3)+\rho_f(2\beta_2-\beta_3) )\mu_0^2
\nn
&+(6P_f(0)(1-\beta_2-\beta_3)+2\rho_f)\mu_0
\nn
&+3P_f(0)(1-2\beta_2)+\rho_f=0\,,
\end{align}
thus there are two branches (the branch C and D) similar to the case of $g$-star.
The branch C approaches a homothetic spacetime as we will see later.

\begin{figure}[tbp]
\centering
\includegraphics[width=8cm,angle=0,clip]{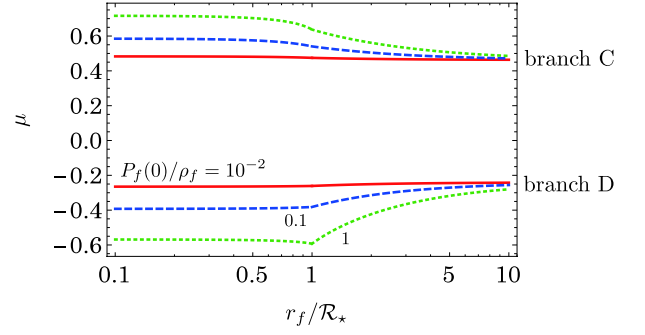}
\caption{The same figure as Fig. \ref{fig_massless_limit}
in the case of $f$-star. 
}
\label{fig_massless_limit2}
\end{figure}

We chose the coupling constants as \eqref{coupling_choice} in Class [I].
The solution in the massless limit is shown in Fig. \ref{fig_massless_limit2}.
For the case of the $f$-star, the wormhole geometry is not found.
Now we solve the basic equations for each branch without the massless limit approximation.

\subsubsection{{\rm Branch C}}
We set
\begin{align}
m_g^2&=m_f^2\,,\quad
\beta_2=-3\,,\quad
\beta_3=3\,, \\
\kappa_f^2\rho_f/m_{\rm eff}^2&=2.5 \times 10^5 \,,\quad
P_f(0)/\rho_f=5\times 10^{-2}\,,
\end{align}
and $\rho_f =$ constant.

\begin{figure}[tbp]
\centering
\includegraphics[width=7cm,angle=0,clip]{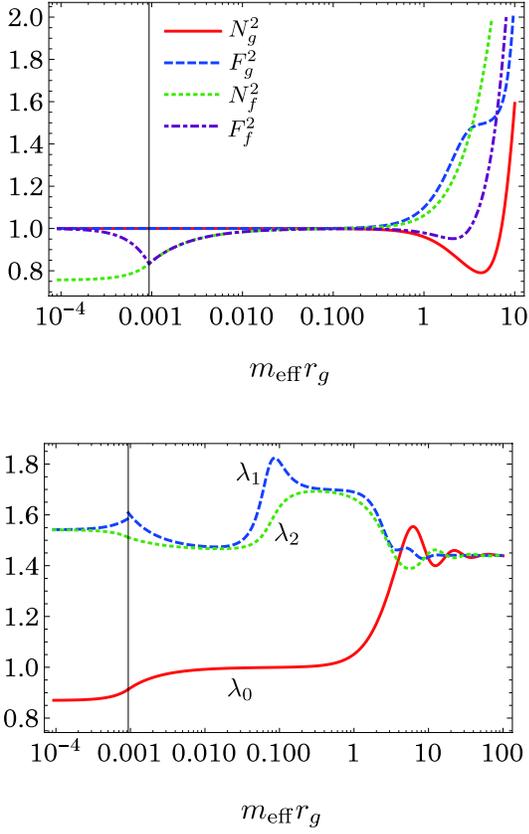}
\caption{
A typical solution for branch C.
We set $P_f/\rho_f=5\times 10^{-2}$.
The shooting parameter is tuned to be $\mu(0)=0.541$.
}
\label{fig_C}
\end{figure}

For the above parameter setting with $K=1$,
an asymptotically AdS solution is found for
$-0.76 m_{\rm eff}^2 \lesssim \Lambda_g(1) \lesssim 0.05m_{\rm eff}^2$.
This solution in the branch C is asymptotically homothetic
because the eigenvalues $\lambda_0,\lambda_1,\lambda_2 $ converges to the same constant 
although its value is not unity
as shown in Fig. \ref{fig_C}.

The reason is as follows:
When we fix parameters $\{m_g,m_f,\Lambda_g,K,\beta_2,\beta_3\}$,
the original coupling constants $\{\kappa_f,b_i\}$ are determined.
Once the original coupling constants are given,
all homothetic solutions given by 
\begin{align}
f_{\mu\nu}=\tilde{K}^2 g_{\mu\nu}
\end{align}
are characterized by the proportional factor $\tilde{K}$ 
which is one of the roots of the quartic equation 
\begin{align}
\Lambda_g(\tilde{K})=\tilde{K}^2\Lambda_f(\tilde{K})\,.
\end{align}
In the range of  $-0.76 m_{\rm eff}^2 \lesssim \Lambda_g(1) \lesssim 0.05m_{\rm eff}^2$,
there are four real roots for $\tilde K$.
For instance, when we set $\Lambda_g(1)=0$,
we find 
\begin{align}
\tilde{K}=-0.604,\,1,\,1.44,\,{\rm and}~3.83\,,
\end{align}
and find four homothetic solutions (one Minkowski, one de Sitter, and two AdS spacetimes).
It turns out that 
the solution we solved approaches $\tilde{K}=1.44$ homothetic spacetime.
Since $\Lambda_g(1.44)<0 $, 
it is the asymptotically AdS spacetime.

Note that 
when we assume $\Lambda_g(1)\lesssim -0.76m_{\rm eff}^2$,
there are only two real roots of $\tilde K$,
e.g,
\begin{align}
\tilde{K}=-0.586,\,{\rm and}~1\,,
\end{align}
for $\Lambda_g(1)=-m_{\rm eff}^2$.
In this case,
 we cannot find a regular solution for the 
 branch C in $\Lambda_g(1)\lesssim -0.76m_{\rm eff}^2$.
In the case of $3m_{\rm eff}^2/2 \gg \Lambda_g(1)\gtrsim 0.05 m_{\rm eff}^2$,
there are four homothetic solutions,
e.g.,
\begin{align}
\tilde{K}=-0.621,\, 1,\, 1.11,\,{\rm and}~ 4.85\,.
\end{align}
for  $\Lambda_g=0.1m_{\rm eff}^2$.
The solution may approach the $\tilde{K}=1.11$ homothetic solution
with $\Lambda_g(1.11)>0$.
However, because of a numerical instability,
we cannot confirm that there is a regular solution approaching 
de Sitter spacetime
for $3m_{\rm eff}^2/2 \gtrsim  \Lambda_g(1)\gtrsim 0.05 m_{\rm eff}^2$.

Finally, we give a comment for the case of $\Lambda_g(1) \gtrsim 3m_{\rm eff}^2/2$.
In this case, the Jacobian $J=dr_f/dr_g$ 
diverges before reaching the cosmological horizon.
Therefore, this solution has the curvature singularity
as discussed in Appendix \ref{sec_wormhole}.

\subsubsection{{\rm Branch D}}
For the branch D, we cannot construct
any regular solution with or without a cosmological constant
by our numerical approach.
Although the solution is regular below the Vainshtein radius,
there is a singularity at a radius near the Compton wavelength of the massive graviton.
Thus we will not discuss the branch D furthermore.

\section{Wormhole-type solution}
\label{sec_wormhole}

In Class [I], as shown in Fig. \ref{fig_massless_limit} (b), we cannot find 
a regular solution 
beyond the critical value of the pressure.
The solution turns to a closed spacetime or 
a wormhole-type spacetime beyond the critical value.

In this appendix, we shall discuss what kind of wormhole type structure is 
obtained in the bigravity theory.
To find a solution with a wormhole-type structure, 
we should integrate the basic equations from the wormhole throat.
As mentioned in the subsection \ref{sec_uniform},
a wormhole throat corresponds to the point of $J=\infty$, where 
the function $J$ is the Jacobian for the radial coordinate transformation 
from $r_g$ to $r_f$.
When we find $J=\infty$ at some radius, such a coordinate transformation is singular.
That is, we cannot define the transformation  $r_g\rightarrow r_f$ at the point.
Similarly, we cannot define the transformation $r_f\rightarrow r_g$  at the point of $J=0$.

When the coordinate transformation $r_f=r_f(r_g)$ is not well-defined (i.e., $J=\infty$) 
at some point, 
we cannot integrate beyond such a singular point as a function of $r_g$.
However, the inverse function $r_g=r_g(r_f)$ is well-defined at $J=\infty$.
As a result, we can solve the equations and find the solution  as a function of $r_f$
by using the radial coordinate $r_f$, i.e.,  the basic equations to be solved are
\begin{align}
\frac{dX}{dr_f}&=J^{-1}\mathcal{F}_X\,,
\\
\frac{dr_g}{dr_f}&=
-\frac{r_f}{(1+\mu)^{2}}\frac{d\mu}{dr_f}+\frac{1}{1+\mu}
\nn
&=J^{-1}\,.
\label{Jacobian2}
\end{align}
Although the point of $J=\infty$ is a curvature singularity as we will see,
we can continue to solve the equations and find the solution 
 beyond such a singularity. 
 
For simplicity, we assume vacuum spacetimes, i.e., 
there is neither $g$-matter nor $f$-matter.
A wormhole throat of $g$-spacetime is given by 
$J^{-1}=dr_g/dr_f=0\,,$
at which
we assume the variables $N_g,F_g,N_f,F_f,\mu$ are finite.
Setting the radial coordinate as $r=r_f$, we find 
the derivatives of $g$-variables are finite  at $J^{-1}=0$ 
because Eqs. \eqref{Fg_eq} and \eqref{Ng_eq} yield
\begin{align}
\frac{dF_g}{dr_f}&\rightarrow -\frac{m_g^2r_f}{2F_f}(1+2\beta_2 \mu+\beta_3\mu^2)\,,\\
\frac{dN_g}{dr_f}&\rightarrow 0\,.
\end{align}
for $J^{-1}\rightarrow 0$.
Furthermore,
Eqs. \eqref{Ff_eq} and \eqref{Nf_eq}
indicate that the derivatives of $f$-variables are also finite at $J^{-1}=0$,
and Eq. \eqref{Jacobian2} indicates $d\mu/dr_f$ is  finite.
Hence, the first derivatives of all variables are finite even at $J^{-1}=0$.
Since the differential equations are first order,
we can solve the equations numerically beyond $J^{-1}=0$ 
by use of the  $r_f$ coordinate.

Since two metric are symmetric in the bigravity theory,
the above argument is also applied to the point of $J=0$,
which is a wormhole throat in $f$-spacetime,
At $J=0$, the coordinate transformation $r_g=r_g(r_f)$ is not well-defined,
but the solution is obtained  as a function of $r_g$
beyond this singularity.

In the case of $\Lambda_g=0$, 
the branch B solution contains a singularity at some radial point.
To find a regular wormhole-type solution, we should introduce a negative cosmological constant.
Here we set the parameters as
\begin{align}
m_g^2&=m_f^2\,,\quad \beta_2=-3\,,\quad
\beta_3=3\,,
\end{align}
and
\begin{align}
\Lambda_g&=-75 m_{\rm eff}^2\quad (\ell_{\rm AdS}=0.2  m_{\rm eff}^{-1}\, )\,.
\label{coupling_AdS}
\end{align}

We first use the $g$-radial coordinate $r_g$.
Suppose that a wormhole throat exists in the $f$-spacetime
(which we call the $f$-throat),
so $J=0$ at a radius $r_g=a_f$
The value of $N_g$ on the throat is arbitrary by the rescaling freedom of the time coordinate,
and the value of $F_g$ gives the gravitational field strength at the throat,
which characterize the property of the wormhole.
Since we have two algebraic equations
at the $f$-throat as
\begin{align}
J|_{r_g=a_f}=0\,,\quad \mathcal{C}|_{r_g=a_f}=0\,,
\label{throat_cond}
\end{align}
where $\mathcal{C}$ is the constraint equation defined by Eq. (\ref{algebraic_eq}),
 when we give the values of $F_g$ and $N_g$ at $r_g=a_f$,
the values $N_f(a_f), F_f(a_f)$ are determined by 
Eqs. \eqref{throat_cond} as functions of $\mu(a_f)$.

\begin{figure}[tbp]
\centering
\includegraphics[width=7cm,angle=0,clip]{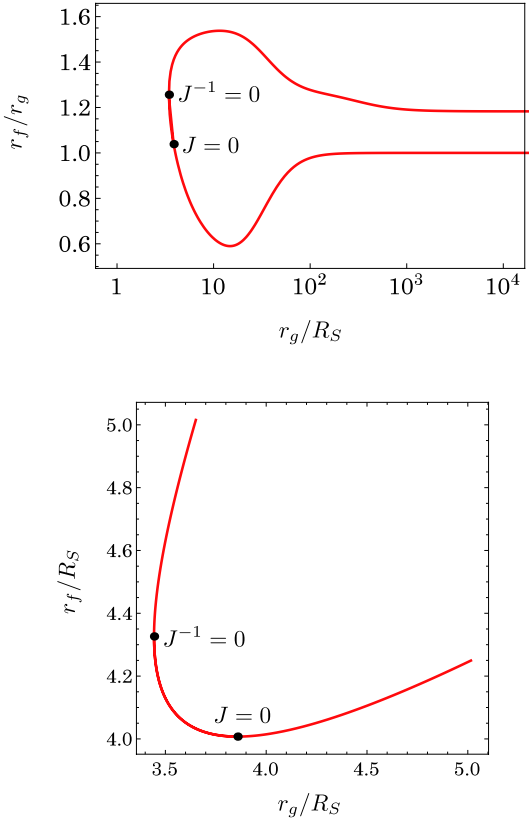}
\caption{
The relation of two radial coordinates $r_g$ and $r_f$.
From the top panel, we find the ratio approaches $r_f/r_g \rightarrow 1$ or 
$r_f/r_g\rightarrow 1.183$ as $r_g \rightarrow \infty$.
$J=0$ and $J^{-1}=0$ correspond to the wormhole throats of $f$-spacetime
 and  of $g$-spacetime, respectively. 
}
\label{wormhole_sol1}
\end{figure}

\begin{figure}[tbp]
\centering
\includegraphics[width=8cm,angle=0,clip]{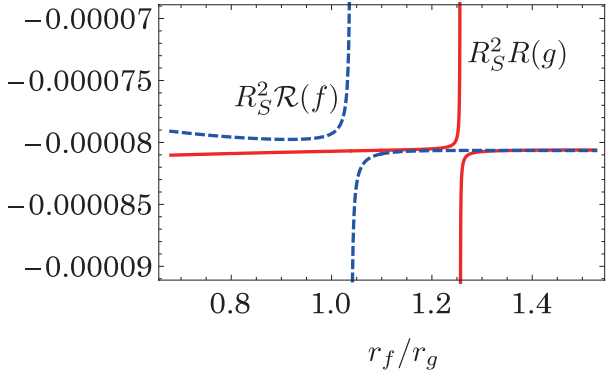}
\caption{
Ricci scalars of $g$-spacetime and $f$-spacetime
which are presented by the red curve and the blue dashed curve, respectively.
The $g$-throat ($J^{-1}=0$) corresponds to $r_f/r_g=1.2563$,
while the $f$-throat ($J=0$) exists at $r_f/r_g=1.03847$.
}
\label{wormhole_sol2}
\end{figure}

We first solve variables outward on the $r_g$ coordinate system,
and find an asymptotically homothetic AdS spacetime 
by tuning the value of $\mu(a_f)$.
Next, we solve variables inward with respect to the $r_g$ coordinate.
When we find the point of $J^{-1}=0$ at a radius $r_g=a_g$, 
which is the wormhole throat in $g$-spacetime (the $g$-throat)
\footnote{The throat condition \eqref{throat_cond} is different 
from the analysis in \cite{Volkov_wormhole},
which paper assumed that two wormhole throats are located at the same spacetime point.
However, we assume, although both spacetimes show wormhole structures,
two wormhole throats are located at the different spacetime points
as shown in Fig. \ref{wormhole_sol1}.}.
we cannot continue to integrate the basic equations numerically on the $r_g$ coordinate.
Then we switch the radial coordinate from $r_g$ to $r_f$,
and solve variables with respect to the $r_f$ coordinate beyond the point of 
$J^{-1}=0$.
Finally we find a global wormhole-type solution, which example is 
given in Figs. \ref{wormhole_sol1}
and \ref{wormhole_sol2} by setting
the graviton mass as $m_{\rm eff}=2\times 10^{-3}  a_f^{-1}$
and by choosing
\begin{align}
N_g(a_f)=F_g(a_f)=0.86070\,.
\end{align}
$\mu(a_f)$ is tuned as
$\mu(a_f)=0.03847$, which gives the asymptotically AdS spacetime.
Here we have introduced a typical length scale of the wormhole $r_S$ by 
\begin{align}
r_S:=2GM_g(\infty)\,,
\end{align}
where we define a mass function $M_g(r)$ by
\begin{align}
F_g^2(r)=1-\frac{2GM_g(r)}{r_g}-\frac{\Lambda_g}{3}r_g^2\,.
\end{align}

Fig. \ref{wormhole_sol1} shows the relation between two radial coordinates.
The top panel gives $r_f/r_g$ in terms of $r_g$ coordinate.
It shows that has $r_f/r_g$ takes two different values at the same radius $r_g$.
One branch $(r_f/r_g\rightarrow 1)$ approaches the homothetic AdS spacetimes,
while another branch $(r_f/r_g \rightarrow 1.183)$ 
approaches the non-homothetic AdS spacetime.
Two different asymptotic structures are connected by the wormhole.
Fig. \ref{wormhole_sol1}
shows that the $g$-throat and the $f$-throat
are located at the different points.

We depict the Ricci curvature scalar of the $f$-metric as well as one of the $g$-metric
in Fig. \ref{wormhole_sol2},
where we have used the variable $r_f/r_g$ to parametrize the radial coordinate,
instead of either $r_g$ or $r_f$,
because either coordinate $r_g$ or $r_f$ is not a single-valued function near the throats.
The $g$-throat ($J^{-1}=0$) is located at $r_f/r_g=1.2563$
and the $f$-throat ($J=0$) is founded at $r_f/r_g=1.03847$.
The Ricci curvature scalar of the $g$-metric  diverges at the $g$-throat.
It is caused by
the divergence of the $\gamma$ energy-momentum tensor at the wormhole throat.
As shown in Fig. \ref{wormhole_sol2},
Ricci scalar 
goes to $+\infty$ as $r_f/r_g \rightarrow  1.2563-\epsilon$,
while it goes to $-\infty$ as $r_f/r_g \rightarrow  1.2563+\epsilon$
with $0<\epsilon \ll 1$.
Note that $f$-spacetime curvature is finite even at the $g$-throat of $J^{-1}=0$.
Only the $g$-spacetime Ricci curvature diverges.
Inversely,
only the $f$-spacetime Ricci scalar diverges at the $f$-throat.
This behaviour is quite similar to the case of the cosmology \cite{with_twin_matter}.

\begin{figure}[tbp]
\centering
\includegraphics[width=7cm,angle=0,clip]{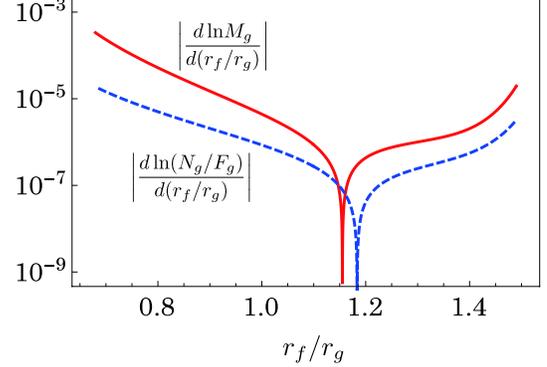}
\caption{The variation rates
of the mass function $M_g$ (red solid) and 
the ratio
$N_g/F_g$ (blue dashed)
as functions of $r_f/r_g$.
}
\label{mass_functions}
\end{figure}

Finally, we discuss the Vainshtein screening. 
Since the $\gamma$ energy-momentum tensor cannot be ignored 
at the throat point,
the Vainshtein screening mechanism is no longer guaranteed.
We may find a deviation from the GR result.
In fact, the geometry of the vacuum spacetime 
turns to a wormhole geometry, 
which does never appear in GR.
To see the differences of the metric functions from GR,
we show the variation rates of the mass function $M_g$ and the ratio $N_g/F_g$
in Fig. \ref{mass_functions}.
In GR, two functions are exactly constant.
In the bigravity, although two functions are not exactly constant,
these are almost constant.
Hence, the metric functions are well-approximated by the Schwarzschild-AdS metric (up to their first derivatives)
although the topology of the solution is different from the Schwarzschild-AdS spacetime.


\end{document}